\documentclass[a4paper,fleqn,usenatbib]{mnras}

\usepackage{amsmath,amssymb}
\usepackage[T1]{fontenc}
\usepackage{ae,aecompl}
\usepackage[utf8]{inputenc}
\usepackage{longtable}
\usepackage{array,booktabs,ragged2e}
\usepackage{graphicx}
\usepackage{subfig}
\usepackage{graphicx}	% Including figure files
\usepackage{verbatim}
\usepackage{amsmath}	% Advanced maths commands
\usepackage{amssymb}	% Extra maths symbols
\usepackage{multirow}
\usepackage{pdflscape}
\usepackage{deluxetable}
\usepackage{subfig}

\title[The Local Void  as traced by LV-SMGPS]{HI Galaxy Signatures in the SARAO MeerKAT Galactic Plane Survey – II. The Local Void and its substructure}

\author[S. Kurapati et al.]{
Sushma Kurapati$^{1}$\thanks{E-mail: sushma@ast.uct.ac.za },
Ren\'ee C. Kraan-Korteweg$^{1}$, 
D.J. Pisano$^{1}$, 
Hao Chen$^{1, 2}$,
\newauthor
Sambatriniaina H. A. Rajohnson$^{1}$,
Nadia Steyn$^{1,3}$,
Bradley Frank$^{4,5,6, 1}$,
Paolo Serra$^{7}$,
\newauthor
Sharmila Goedhart$^{5}$, 
Fernando Camilo$^{5}$
\\ \\
$^{1}$ Department of Astronomy, University of Cape Town, Private Bag X3, Rondebosch 7701, South Africa\\
$^{2}$ Research Center for Intelligent Computing Platforms, Zhejiang Laboratory, Hangzhou 311100, China \\
$^{3}$International Centre for Radio Astronomy Research (ICRAR), The University of Western Australia, 35 Stirling Highway, Australia \\
$^{4}$UK Astronomy Technology Centre, Royal Observatory Edinburgh, Blackford Hill, Edinburgh EH9 3HJ, UK \\
$^{5}$South African Radio Astronomy Observatory, 2 Fir Street, Observatory, 7925, South Africa \\
$^{6}$The Inter-University Institute for Data Intensive Astronomy (IDIA), and University of Cape Town, Private Bag X3, Rondebosch, 7701, South Africa \\
$^{7}$INAF – Osservatorio Astronomico di Cagliari, Via della Scienza 5, 09047, Selargius, CA, Italy
}

\date{Accepted XXX. Received YYY; in original form ZZZ}

\pubyear{2023}

\begin{document}
\label{firstpage}
\pagerange{\pageref{firstpage}--\pageref{lastpage}}
\maketitle

\newcommand{\HI}{\rm H{\sc i }}
\newcommand{\MB}{\ensuremath{\rm M_B}}
\newcommand{\mb}{\ensuremath{\rm M_{bar}}}
\newcommand{\mhi}{\ensuremath{\rm M_{HI}}}
\newcommand{\mg}{\ensuremath{\rm M_{g}}}
\newcommand{\ms}{\ensuremath{\rm M_{s}}}
\newcommand{\msun}{\ensuremath{\rm M_{\odot}}}
\newcommand{\fatm}{\ensuremath{\rm f_{atm}}}
\newcommand{\jb}{\ensuremath{\rm j_{bar}}}
\newcommand{\js}{\ensuremath{\rm j_s}}
\newcommand{\jg}{\ensuremath{\rm j_{g}}}
\newcommand{\kms}{\ensuremath{\rm km \, s^{-1}}}
\newcommand{\vhel}{\ensuremath{ V_{\rm hel}}}
\newcommand{\kpc}{\ensuremath{\rm kpc}}
\newcommand{\Jykms}{\ensuremath{ \rm Jy \, km \, s^{-1} }}
\newcommand{\mJybeam}{\ensuremath{ \rm mJy \, beam^{-1} }}
\newcommand{\LCDM}{\ensuremath{\Lambda{\rm CDM}}}
\newcommand{\aips}{{\sc aips }}
\newcommand{\gipsy}{{\sc gipsy }}
\newcommand{\fat}{{\sc fat }}
% Abstract of the paper
\begin{abstract}
The Local Void is one of the nearest large voids, located at a distance of 23 Mpc. It lies largely behind the Galactic Bulge and is therefore extremely difficult to observe. We use \HI\ 21 cm emission observations from the SARAO MeerKAT Galactic Plane Survey (SMGPS) to study the Local Void and its surroundings over the Galactic longitude range 329$^{\circ}< \ell <$ 55$^{\circ}$, Galactic latitude $|b| <$ 1.5$^{\circ}$, and redshift $cz <$ 7500 \kms. We have detected 291 galaxies to median rms sensitivity of 0.44 mJy per beam per 44 \kms \ channel. We find 17 galaxies deep inside the Void, 96 at the border of the Void, while the remaining 178 galaxies are in average density environments. The extent of the Void is $\sim$ 58 Mpc. It is severely under-dense for the longitude range 350$^{\circ}< \ell <$ 35$^{\circ}$ up to redshift $z <$ 4500 \kms. The galaxies in the Void tend to have \HI masses that are lower (by approximately 0.25 dex) than their average density counterparts. We find several potential candidates for small groups of galaxies, of which two groups (with 3 members and 5 members) in the Void show signs of filamentary substructure within the Void.

\end{abstract}

% Select between one and six entries from the list of approved keywords.
% Don't make up new ones.

%%%%%%%%%%%%%%%%%%%%%%%%%%%%%%%%%%%%%%%%%%%%%

%%%%%%%%%%%%%%%%% BODY OF PAPER %%%%%%%%%%%%%%%%%%
\section{Introduction}
Galaxies are distributed in a `cosmic web' consisting of voids, interconnected filaments, walls, and nodes surrounding vast low-density regions \citep[e.g.][]{jones09, tempel14}. Voids form prominent features of this `cosmic web' \citep{vande11}. They are extended regions  with sizes in the range of 20 -- 50 h$^{-1}$ Mpc that are largely devoid of galaxies and occupy a major fraction of the volume of the universe. The voids provide a unique opportunity to study galaxies at an earlier stage of their evolution because galaxies in voids typically evolve at later times than those in the dense regions \citep[e.g.][]{aragon13}. For example, extremely low metallicity dwarf galaxies have been found in voids, suggestive of a slow evolution \citep[e.g.][]{pustilnik06, pustilnik11}. The void galaxies are also largely unaffected by complex environmental processes that modify galaxies in higher density environments and thereby retain important clues regarding their formation and evolution. The voids are expected to have substructure consisting of sub-voids, tendrils, filaments, walls, and nodes because the structure formation in voids  is similar to structure formation in a  low density universe \citep{goldberg04, aragon13, alpaslan14}. Numerical simulations suggest that the galaxies in voids may preferentially lie within these filamentary void substructures \citep{sheth04, vande11, aragon13, rieder13}. Hence, it is interesting to explore whether the intricate substructure expected from hierarchical structure formation processes can be traced observationally. 
%accretion

Observations of void galaxies have shown that they are statistically bluer, fainter, hence of later morphological type, and have higher specific star formation rates compared to their counterparts in average density environments \citep{rojas04, rojas05, hoyle05, blanton05, croton05, park07, croton08, hoyle12, ricciardelli14, moorman14, moorman15}. However, controversy persists in the literature as to whether the void galaxies have different intrinsic properties compared to similar galaxies in average density regions \citep[e.g.][]{patiri06, park07, von08, kreckel12, moorman16, kurapati18, kurapati20a, jian22}. Voids contain a population of galaxies that are relatively gas rich, many of which present evidence for ongoing gas accretion, interactions with small companions and filamentary alignments \citep[e.g.][]{ekta08, kreckel11, kreckel12, beygu13, chengalur13, chengalur17, kurapati20}. Numerical simulations predict that low density regions at $z \sim$ 0 are dominated by cold gas accretion \citep[e.g.][]{keres05, keres09, Dekel13}. For example, \citet{stanonik09} found a an extremely extended and massive H{\sc i} polar disk galaxy providing a convincing case for cold mode accretion. Voids are therefore uniquely promising places to search for evidence of ongoing gas accretion.

At a distance of $\sim$ 23 Mpc \citep{tully87, tully08}, the `Local Void' (LV) is the nearest void in the local universe. Studies of cosmic flows in the Local Volume have demonstrated motions of galaxies away from the LV \citep[e.g.][]{tully08, karachentsev15, rizzi17, shaya17, anand18, tully19}.   Studying the size, census, and  emptiness of the LV is critical to understand the distribution of mass responsible for the motion of galaxies. In addition, the proximity of the LV allows us to study its galaxies in detail down to very faint dwarf galaxies. 

However, the LV has been difficult to study since a major fraction lies behind the Galactic Bulge of our own Milky Way. The high foreground extinction and the high stellar density mask galaxies at optical/infrared wavelengths, which impacts a thorough study of the LV despite its proximity. \HI 21-cm line observations is one of the few methods that allows an investigation of its galaxy population through the thickest Galactic obscuration. The Parkes single dish multibeam HI surveys such as HIPASS \citep{meyer04, wong06} and HIZOA \citep[][Kraan-Korteweg et al. in prep]{donley05, kraan08, staveley16} confirmed the LV to be severely underdense. While these shallow surveys uncovered some galaxies in the Local Void, we still do not have a good census of the Void though.

 We therefore decided to make use of the data that was taken as part of the SARAO MeerKAT Galactic Plane Survey (SMGPS; Goedhart et al., in prep). The sensitivity and depth of the SMGPS provides us with a valuable data set to learn more about the void population. In this paper, we present SMGPS H{\sc i } observations of the LV and its surroundings in the Zone of Avoidance (ZOA) over the Galactic longitude range $329^{\circ}< \ell < 55^{\circ}$, Galactic latitude $|b| < 1.5^{\circ}$, and redshift $z < 7500 \ \kms$, which is henceforth referred to as the LV--SMGPS survey. This is the largest blind wide-area interferometric H{\sc i} survey of void galaxies. \citet{Weinberg91} conducted the first blind interferometric H{\sc i} survey in a void situated in the foreground of the Perseus-Pisces supercluster. However, no galaxies were detected in the void during their survey. Other previous high-resolution studies on void galaxies were based on targeted observations around the optically selected galaxies, and hence, are biased by the selection criterion \citep[e.g.][]{szomoru96, kreckel12, kurapati20}.

 %The rest of the paper is organized as follows. In \S \ref{sec:observations}, we describe the observations, data reduction, and source finding strategies. We present the distribution of galaxies in and around the Local Void and discuss   \S \ref{sec:results}

\section{MeerKAT Observations $\&$ data reduction}
\label{sec:observations}

\begin{figure*}
    \centering
    \includegraphics[width=1.0\linewidth]{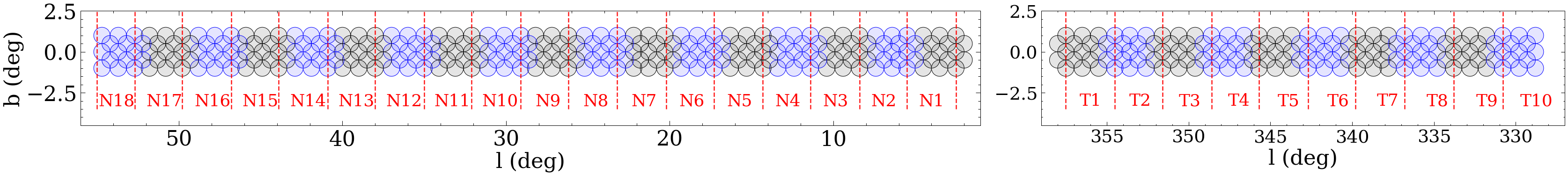}
    \caption{Mosaic configuration of the LV--SMGPS of 28 mosaics of 22 fields each with Galactic longitude, 2$^{\circ}$ $<\ell <$ 55$^{\circ}$  (left: N1- N18) and 329$^{\circ}$ $<\ell <$ 359$^{\circ}$ (right: T1-T10) and $-$1.5$^{\circ}$ $< b <$ 1.5$^{\circ}$. Each circle represents a MeerKAT pointing and our region of interest corresponds to a total of 440 MeerKAT pointings.   }
    \label{fig:mosaic}
\end{figure*}

\subsection{The SARAO MeerKAT Galactic Plane Survey}
SMGPS survey is a systematic \HI survey that covers 248$^{\circ}$ $< \ell <$ 60$^{\circ}$ and $|b|$ $<$ 1.5$^{\circ}$. The survey didn't include the Galactic centre region, 358$^{\circ}$ $< \ell <$ 2$^{\circ}$ which was covered by \citet{Heywood22}. The main scientific goals of the SMGPS are the studies of the Galaxy itself and its objects.  However, we can extract the 21 cm spectral line observations to explore the large scale structure through their H{\sc i} emission and trace hidden structures such as the Great Attractor (GA), the Vela supercluster (VSCL), and the Local Void (LV), the primary focus here. This paper is the second in a series leveraging H{\sc i} data from the SMGPS survey. The first paper explores the Great Attractor (Steyn et al. 2023, under review in MNRAS), while the third paper is on the Vela supercluster (Rajonson et al. in prep),  which together improve our understanding of the dynamics in the local universe.

\subsection{Data reduction}
The observations were carried out with the full MeerKAT array in the L-band (856--1712 MHz)  using the MeerKAT SKARAB 4k-correlator, which gives a frequency resolution of $\sim$ 209 kHz that corresponds to a rest frame velocity resolution $\sim$ 44.1 \kms at $z = 0 $. Each MeerKAT pointing has an effective integration of $\sim$ 1 hour, but the observations  were done in small sub-integrations spread over $\sim$ 10 hours to retain excellent u-v coverage. 

The LV--SMGPS raw data were downloaded from the SARAO Archive for the range 1308–1430 MHz enabling us to detect galaxies out to $\sim$ 25,000 \kms. We use the Containerised Automated Radio Astronomy Calibration ({\tt CARACal}) pipeline \citep{jozsa20} to reduce the raw data.  {\tt CARACal} is based on {\tt STIMELA}, a radio interferometry scripting framework based on container technologies and Python \citep{makhathini18}.  It uses many publicly available radio interferometry software packages such as {\tt CASA}, {\tt Cubical}, {\tt WSClean} and {\tt Montage}, etc. The pipeline carried out all the standard data reduction tasks including flagging, cross-calibration, self-calibration, continuum subtraction, and spectral line imaging.
We perform the standard cross-calibration procedures such as delay, bandpass and gain calibrations, and later use the self-calibration of the continuum to further improve the quality of the calibration. We use {\tt Cubical} \citep{Kenyon18} to carry out two rounds of self calibration. Further, the continuum sky model that was generated after self-calibration is subtracted from the visibilities.

For the \HI\-imaging of the individual fields, we use WSClean with robust=0 weighting and a taper of 15$''$. This gives us dirty H{\sc i} cubes with a nearly Gaussian beam  size of  $30'' \times 27''$  ($\pm 2 ''$), which is sufficient for our science goals.  The H{\sc i} data cubes were also made at lower resolutions using taper 30$''$ and taper 60$''$ to obtain maps with beam sizes closer to FWHM 45$''$ and 67 $''$, ensuring that we do not miss out on nearby very extended low column density sources. A more detailed description of the data reduction strategy and data quality assessment is presented in Rajohnson et al., in prep.

In this paper, we present the data for the region in and around the Galactic Bulge, 329$^{\circ}$ $<$ $\ell$ $<$ 358$^{\circ}$ and 2$^{\circ}$ $<\ell <$ 55$^{\circ}$ and $-$1.5$^{\circ}$ $< b <$ 1.5$^{\circ}$, which corresponds to a total of 440 MeerKAT64 pointings.  Individual data cubes were primary-beam corrected and are stitched together to create mosaic cubes to maximize the sensitivity as shown in Fig. \ref{fig:mosaic}. We use the CARACAL pipeline to construct 28 \HI mosaic cubes, each mosaic cube consisting of 22 contiguous primary-beam corrected \HI data cubes of partly overlapping pointings, covering an area of $4^{\circ}$ $\times$ $3^{\circ}$. The mosaics were offset by $\Delta \ell$ $=$ 3$^{\circ}$  to detect extended sources on the borders of mosaics and to allow for independent assessment of the quality of the data products in the overlapping regions. 
\begin{figure}
    \centering
    \includegraphics[width=0.85\linewidth]{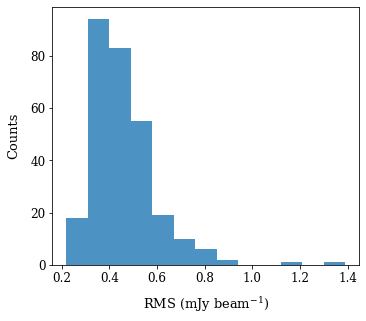}
    \caption{Histogram of RMS noise measured around each detection. The median RMS noise is 0.42 mJy beam$^{-1}$ per 44.1 \kms channel.}
    \label{fig:noise}
\end{figure}

\newcolumntype{R}[1]{>{\RaggedLeft\arraybackslash}p{#1}}
\newcolumntype{L}[1]{>{\RaggedRight\arraybackslash}p{#1}}
\newcolumntype{C}[1]{>{\Centering\arraybackslash}p{#1}}
\begin{table*}
\caption{An extract of the LV--SMGPS catalogue listing \HI parameters of galaxies. This table is published  in its  entirety online in the machine readable format}
%\begin{tabular}{p{2.12cm}p{0.6cm}p{0.5cm}p{0.7cm}p{1.4cm}p{1.4cm}p{0.55cm}p{0.55cm}p{0.95cm}p{0.95cm}p{0.95cm}p{0.95cm}p{0.4cm}p{0.4cm}}
\begin{tabular}{L{2.1cm}L{0.60cm}R{0.70cm}R{0.65cm}R{0.85cm}C{0.85cm}C{0.65cm}C{1.20cm}C{0.65cm}R{0.70cm}R{0.60cm}R{0.9cm}C{0.5cm}L{1.0cm}}
\hline
Name & Mosaic & $\ell$ & $b$ & $S_{\rm int}$ & $\sigma_{Sint}$ & $S_{\rm peak}$ & RMS & \vhel & $w_{\rm 50}$ & $w_{\rm 20}$  & log(M$_{HI}$) & Flag & Note \\
SMGPS-HI- &  & deg & deg & \Jykms & \Jykms & Jy & \mJybeam & \kms & \kms & \kms & log(M$_{\odot}$) & & \\
(1) & (2) & (3) & (4) & (5) & (6) & (7) & (8) & (9) & (10) & (11) & (12) & (13) & (14) \\
\hline
J175227-262618 & N01 & 2.92 & -0.01 & 0.39 & 0.21 & 0.007 & 0.46 & 6024 & 204 & 96 & 8.85 & 1 & \\
J175200-260838 & N01 & 3.12 & 0.23 & 2.62 & 0.33 & 0.010 & 0.59 & 7115 & 326 & 280 & 9.82 & 1 & \\
J175024-254605 & N01 & 3.26 & 0.73 & 2.04 & 0.27 & 0.018 & 0.54 & 6162 & 217 & 188 & 9.59 & 1 & \\
J174806-252644 & N01 & 3.27 & 1.34 & 1.04 & 0.38 & 0.013 & 0.71 & 6024 & 139 & 139 & 9.28 & 1 & \\
J175321-254318 & N01 & 3.64 & 0.18 & 1.75 & 0.25 & 0.026 & 0.54 & 6284 & 107 & 60 & 9.54 & 1 & \\
J175537-254553 & N01 & 3.86 & -0.28 & 3.24 & 0.37 & 0.020 & 0.68 & 7134 & 414 & 381 & 9.92 & 1 & \\
J175725-253712 & N01 & 4.19 & -0.56 & 1.13 & 0.18 & 0.010 & 0.49 & 5886 & 195 & 132 & 9.29 & 1 & \\
J175857-252902 & N01 & 4.48 & -0.79 & 5.46 & 0.88 & 0.030 & 0.56 & 6886 & 276 & 260 & 10.11 & 1 & \\
J175618-250448 & N01 & 4.53 & -0.07 & 4.23 & 0.92 & 0.022 & 0.68 & 7682 & 354 & 312 & 10.10 & 1 & \\
J175258-242216 & N01 & 4.76 & 0.94 & 4.76 & 0.20 & 0.022 & 0.55 & 6459 & 266 & 224 & 10.00 & 1 & \\
J175320-241850 & N01 & 4.85 & 0.90 & 2.30 & 0.12 & 0.018 & 0.45 & 6477 & 186 & 163 & 9.68 & 1 & \\
J175325-241645 & N01 & 4.89 & 0.90 & 1.81 & 0.12 & 0.014 & 0.48 & 6508 & 163 & 139 & 9.58 & 1 & \\
J175330-241811 & N01 & 4.88 & 0.87 & 1.46 & 0.23 & 0.010 & 0.45 & 6325 & 186 & 139 & 9.46 & 1 & \\
\hline
\label{table:example}
\end{tabular}
\end{table*}
\subsection{HI detections}
We use the Source Finding Application SoFiA pipeline  \citep{serra15,Westmeier21} which is designed for automatic H{\sc i} source finding  in interferometric spectral line 
data. SoFiA2 (\url{https://github.com/SoFiA-Admin/SoFiA-2}) is applied to detect  H{\sc i} sources in the redshift range $300 \ \kms < \vhel < 7500 \ \kms$ which 
covers the full depth of the LV with its borders but excludes the channels containing Milky Way emission. We do not consider the edges of the mosaic images, where sensitivity is reduced, and the SoFiA search area is limited to $3.5^{\circ}$ $\times$ $3^{\circ}$ with a remaining overlap of 0.25$^{\circ}$ on each side of the mosaic.
Within SoFiA, we remove noise variations in the mosaic cubes by using the noise images that were obtained as part of mosaicking algorithm. We use the `smooth+clip' algorithm which smooths the cube spatially and spectrally with multiple user-defined Gaussian and boxcar kernels respectively. We apply a threshold of 5 $\sigma$ using 4 Gaussian smoothing kernels that resulted in angular resolutions of no smoothing, 30$''$, 60$''$, and 90$''$, in addition to three spectral boxcar kernels of no smoothing, 133, and 308 \kms. We merge the detected voxels into individual sources by using a merging radius of a beam along both the RA and Dec axes, and 2 channels (with channel width $\sim$ 44.1 \kms) in velocity. 

Each of the SoFiA runs yields a list of real and false positive detections. It is not possible to assess the reliability of sources by searching for a plausible optical counterpart within the SoFiA-2 mask since we do not have any multi-wavelength data this close to the Galactic plane. We therefore inspect the channel maps, moment maps, and spectra of all possible detections obtained using SoFiA to assess the reliability of galaxy candidates. We derive the moment maps of all the possible detections by smoothing the cubelets centred on each detection to a circular beam of 35'' $\times$ 35'' and clipping it at 3$\sigma$ threshold, where $\sigma$ is the average RMS noise around each detection \citep[e.g.][]{ponomareva21}. Moment maps were then derived by applying the resulting mask to the original resolution unsmoothed cubelet to account for a low column density diffuse \HI emission. This mask was also used to derive the spectrum from the unsmoothed cube for each source. Further, we check for velocity continuity of \HI emission, expected \HI size for a given \HI mass, significance of \HI emission in the spectrum, etc to classify each source as either a definite detection or a false detection. This leads to a total of 291 definite detections. 

To determine the RMS noise levels around each detection, we take the average RMS from four emission-free regions around the detection. The histogram of the RMS noise is shown in Fig. \ref{fig:noise}. The RMS noise per channel varies from 0.22 to 1.32 \mJybeam\ with a median noise level of  RMS $=$ 0.42 \mJybeam.   We show masked moment-0 (left panel), masked moment-1 map (middle panel), and the global H{\sc i} profile (right panel) of some of the definite detections in Figure \ref{fig:mom1} in Appendix. The moment maps and spectra for all the detections are available online. The galaxy ID (Jhhmmss $\pm$ ddmmss) for each galaxy is shown on the top-left of moment 0 map. The synthesis beam FWHM is indicated by an ellipse in both the moment 0 and moment 1 maps. The H{\sc i} global profile is shown by solid black line, the smoothed profile by a black dashed line, the solid red line indicates the heliocentric velocity.

%using the following procedure \citep[e.g.][]{ponomareva21}. The cubelets centred on each detection were smoothed to a circular beam of 35'' $\times$ 35'' and clipped at 3$\sigma$ threshold, where $\sigma$ is the measured average noise around each detection. 

We calculate the integrated flux from the SoFiA output as well as the spectrum. The fluxes measured with the two methods agree within $\sim$ 10 $\%$. To calculate the error of the integrated flux, we project the 3D source mask on to eight emission free regions (four regions at vsys $-$ 220 \kms and other four at vsys $+$ 220 \kms )  surrounding the detection in voxels and measure the signal in each of the 8 projected masks. The uncertainty in the integrated flux is then defined as the RMS scatter from the flux measurements in the eight projected masks \citep[e.g.][]{ramatsoku16, ponomareva21}. Further, we calculate the H{\sc i} mass using the standard formula,  $M_{\rm HI} = 2.356\times 10^{5} D^{2} S_{\rm int} \ {\rm M}_{\rm \odot}$, where $D$ is the distance in Mpc, $S_{\rm int}$ is the integrated flux density in Jy km  s$^{-1}$. The \HI redshift was extracted from the SoFiA output and the luminosity distance was calculated for each source based on its \HI redshift and assuming the cosmological values of $\Omega_{M} =$ 0.3, $\Omega_{\Lambda} =$ 0.7, and H$_{0} = $  70 \kms \ Mpc$^{-1}$ from the standard $\Lambda$CDM cosmology.

\subsubsection{LV-SMGPS Catalogue}
An extract of the LV--SMGPS catalogue listing  the \HI parameters for the detected galaxies is shown in Table \ref{table:example}. The full catalogue is given in table \ref{table:app1} in the Appendix.

The columns are as follows: \\
{\it Column 1} - Name of the galaxy: The format (SMGPS--HI) J{\small HHMMSS-DDMMSS} represents the rounded values of RA and Declination  \\
{\it Column 2} -  Mosaic (Txx or Nxx from Fig. \ref{fig:mosaic}) in which the galaxy is detected \\
{\it Column 3 and 4} -  Galactic coordinates, $\ell$ and $b$ in deg \\
{\it Column 5 and 6 } -  Integrated flux, $S_{\rm int}$ along with its corresponding error $\sigma_{S_{\rm int}}$ in \Jykms  \\
{\it Column 7} -  Peak flux, $S_{\rm peak}$ in Jy \\
{\it Column 8} -  Measured local RMS around the detection in \mJybeam\  per 44.1 \kms channel \\
{\it Column 9} - Optical heliocentric velocity, $V_{\rm hel}$ in \kms \\
{\it Column 10 and 11} - Line widths $w_{\rm 50}$ and $w_{\rm 20}$ in \kms. These values represent the measured linewidths at 50 $\%$ and 20 $\%$ of the peak flux, respectively \\
{\it Column 12} -  Logarithm of the \HI mass, log $M_{\rm HI}$ in log (\msun) \\
{\it Column 13} - Flag (category 1 indicates a definite detection and category 2 indicates a possible detection) \\
{\it Column 14} -  Notes on galaxies. It describes whether a galaxy is interacting, has a companion, or any other relevant information.

\begin{table*}
\caption{Comparison of HIZOA and LV-SMGPS parameters for the HIZOA sources }
\begin{tabular}{lrrrrrrrr}
\hline
HIZOA ID & SMGPS ID & $\ell$ & $b$ & \vhel & $S_{\rm int, HIZOA}$ & $S_{\rm int, SMGPS}$  & log $M_{\rm HI}$ & log $M_{\rm HI}$ \\
         &        & deg & deg & \kms & \Jykms & \Jykms &  log (\msun) &  log (\msun) \\
  (1)    & (2)    & (3) & (4) & (5) & (6) & (7) &  (8) &  (9) \\
\hline
J1603$-$53 & J160305$-$531800 & 329.26 & -0.52 & 5797 & 6.68 $\pm$ 1.15 & 3.93 $\pm$	0.19 & 10.04 & 9.82 \\
J1607$-$53 & J160730$-$534932 & 329.40 & -1.35 & 5497 & 7.75 $\pm$  1.21 & 7.86 $\pm$	0.57 & 10.07 & 10.07 \\
J1605$-$51 & J160544$-$511338 & 330.94 & 0.76 & 5723 & 10.63 $\pm$ 2.85 & 4.51 $\pm$	0.28 & 10.24 & 9.86 \\
J1612$-$52 & J161245$-$524536 & 330.70 & -1.10 & 5684 & 2.45 $\pm$ 0.90 & 2.97 $\pm$	0.16 & 9.59 & 9.68 \\
J1607$-$51 & J160749$-$510452 & 331.28 & 0.65 & 6041 & 8.65 $\pm$  1.51 & 8.77 $\pm$	0.23 & 10.19 & 10.20 \\
J1609$-$50 & J161007$-$501128 & 332.15 & 1.06 & 6251 & 5.18 $\pm$  2.01 & 5.33 $\pm$	0.34 & 10.0 & 10.01 \\
J1617$-$49 & J161743$-$493216 & 333.49 & 0.69 & 2409 & 10.59 $\pm$  1.76 & 11.35 $\pm$	0.22 & 9.48 & 9.51 \\
J1627$-$47 & J162704$-$473758 & 335.93 & 0.94 & 7056 & 9.26 $\pm$  1.28 & 7.53 $\pm$	0.48 & 10.36 & 10.27 \\
J1626$-$47 & J162702$-$471646 & 336.18 & 1.19 & 7016 & 6.60 $\pm$  1.17 & 4.35 $\pm$	0.37 & 10.20 & 10.03 \\
J1640$-$45 & J164039$-$445933 & 339.46 & 1.01 & 4625 & 5.72 $\pm$  1.55 & 3.69 $\pm$	0.24 & 9.78 & 9.59 \\
J1656$-$44 & J165639$-$440511 & 341.99 & -0.59 & 4903 & 10.04 $\pm$  2.45 & 11.95 $\pm$	0.46 & 10.07 & 10.15 \\
J1657$-$45 & J165744$-$451328 & 341.22 & -1.45 & 6088 & 4.01 $\pm$  1.22 & 2.87 $\pm$	0.38 & 9.87 & 9.72 \\
J1704$-$41 & J170448$-$414022 & 344.81 & -0.28 & 5525 & 5.44 $\pm$  1.40 & 6.63 $\pm$	0.33 & 9.91 & 10.00 \\
J1705$-$40 & J170527$-$405659 & 345.46 & 0.06 & 3920 & 4.61 $\pm$  1.58 & 5.98 $\pm$	0.36 & 9.54 & 9.65 \\
J1713$-$40 & J171314$-$405045 & 346.42 & -1.06 & 6046 & 4.07 $\pm$  1.19 & 3.99 $\pm$	0.23 & 9.87 & 9.86 \\
J1719$-$37 & J171906$-$373147 & 349.78 & -0.06 & 3411 & 3.05 $\pm$  0.90 & 6.48 $\pm$	0.46 & 9.24 & 9.57 \\
J1758$-$21 & J175830$-$215009 & 7.59 & 1.12 & 6909 & 5.87 $\pm$  1.35 & 5.29 $\pm$	0.40 & 10.14 & 10.10 \\
J1808$-$21 & J180808$-$213547 & 8.90 & -0.71 & 1466 & 3.25 $\pm$  1.00 & 2.22 $\pm$	0.28 & 8.53 & 8.37 \\
J1904+03A & J190412+030211 & 37.09 & -1.45 & 3308 & 5.60 $\pm$ 0.84 & 3.93 $\pm$	0.32 & 9.47 & 9.33 \\
J1901+06 & J190135+065133 & 40.19 & 0.88 & 2940 & 17.20 $\pm$ 2.58 & 14.55 $\pm$	0.22 & 9.86 & 9.79 \\
J1908+05 & J190824+055946 & 40.20 & -1.02 & 4565 & 5.97 $\pm$ 0.89 & 5.24 $\pm$	0.13 & 9.79 & 9.74 \\
J1906+07 & J190640+073452 & 41.41 & 0.09 & 3097 & 5.16 $\pm$  0.77 & 3.42 $\pm$	0.41 & 9.38 & 9.21 \\
J1914+10 & J191500+101718 & 44.76 & -0.48 & 637 & 20.60 $\pm$ 3.09 & 20.67 $\pm$	0.55 & 8.63 & 8.60 \\
J1912+13 & J191236+132340 & 47.24 & 1.48 & 2764 & 7.09 $\pm$ 1.06 & 9.15 $\pm$	1.07 & 9.43 & 9.54 \\
J1919+14 & J191957+140453 & 48.68 & 0.22 & 2797 & 13.70$\pm$  2.05 & 12.97 $\pm$	1.27 & 9.72 & 9.69 \\
J1921+14 & J192135+145324 & 49.58 & 0.25 & 4075 & 6.56 $\pm$ 0.98 & 8.28 $\pm$	0.40 & 9.73 & 9.83 \\
J1918+16 & J191846+161005 & 50.39 & 1.45 & 6630 & 7.08 $\pm$ 1.06 & 4.71 $\pm$	0.30 & 10.19 & 10.01 \\
\hline
\label{table:HIZOA}
\end{tabular}
\end{table*}
\begin{figure}
    \centering
    \includegraphics[width=1.06\linewidth]{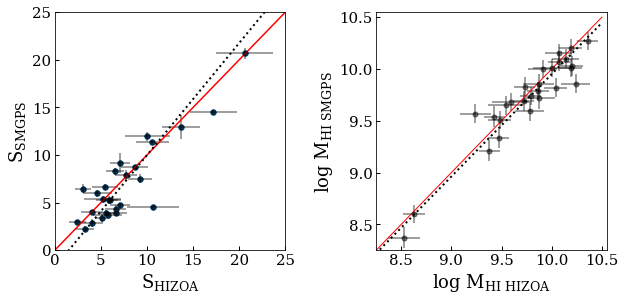}
    \caption{Comparison of integrated flux in \Jykms (left panel) and HI Mass in log(\msun)  (right panel) between LV--SMGPS and HIZOA. The red solid line represents x=y line and the black dotted line represents the linear regression. }
    \label{fig:HIZOA_GPS}
\end{figure}

%\begin{figure}
%    \centering
%    \includegraphics[width=0.51\linewidth]{HIZOA_GPS_flux.png}
%    \includegraphics[width=0.47\linewidth]{MHI_HIZOA_GPS.png}
%    \caption{Left panel: Flux comparison of MeerKAT-GPS and the HIZOA observations, right panel: HI mass comparison of MeerKAT-GPS versus HIZOA observations. The red solid line represents x=y line and the black dotted line represents the linear regression. }
%    \label{fig:HIZOA_GPS}
%\end{figure}

\subsubsection{HI data assessment}

We verified the measured \HI fluxes and \HI masses by comparing them to the detections from the Parkes HI Zone of Avoidance (HIZOA) survey \citep{staveley16} and the Northern Extension of HIZOA survey \citep{donley05}. The HIZOA survey is a `blind’ wide area systematic \HI survey of the southern Zone of Avoidance  using the Parkes single dish telescope equipped with the multi-beam receiver ($|b| <$ 5$^{\circ}$, RMS $\sim 6 \ \mJybeam$ ). There are 32 galaxies in this region that have HIZOA counterparts. All HIZOA galaxies are recovered in our analysis. Five of the galaxies were found to be resolved into two or more galaxies by the LV--SMGPS observations: HIZOA J1753-24A is conformed of 4 galaxies (see `Group B3' in Fig. \ref{fig:Void_Border_groups}), while  HIZOA J1811-21, HIZOA J1653-44, HIZOA J1721-37, and HIZOA J1716-35 are resolved into pairs of galaxies. 

Table \ref{table:HIZOA} lists the parameters of galaxies from both the HIZOA survey and the LV--SMGPS survey. For the flux verification, we only use HIZOA galaxies that have one counterpart in LV--SMGPS survey.
%and the MKT-GPS fluxes are consistent with HIZOA fluxes {\it within errorbars for 23 of 27 galaxies}. 
This is illustrated in Fig. \ref{fig:HIZOA_GPS}. The left panel of  Fig. \ref{fig:HIZOA_GPS} displays the HIZOA flux versus LV--SMGPS flux. The red solid line indicates the x=y line while the black dotted line indicates the linear regression. The slope for the linear regression is 1.18 $\pm$ 0.13, which is consistent with slope 1 within $\sim$ 1.4 $\sigma$ indicating that there is no systematic deviation between HIZOA and LV--SMGPS fluxes. The right panel in Fig. \ref{fig:HIZOA_GPS} shows the \HI mass derived from the HIZOA flux versus the \HI mass derived from the LV--SMGPS flux. We estimate the \HI mass of HIZOA galaxies based on the luminosity distances that were calculated based on their heliocentric velocities to make it consistent with LV--SMGPS \HI mass estimates. The black dotted line shows the linear regression with a slope of 0.989 $\pm$ 0.07, which indicates that the \HI masses measured from the HIZOA survey match well to those from the LV--SMGPS survey. 

\begin{figure*}
    \centering
    \includegraphics[width=0.95\linewidth]{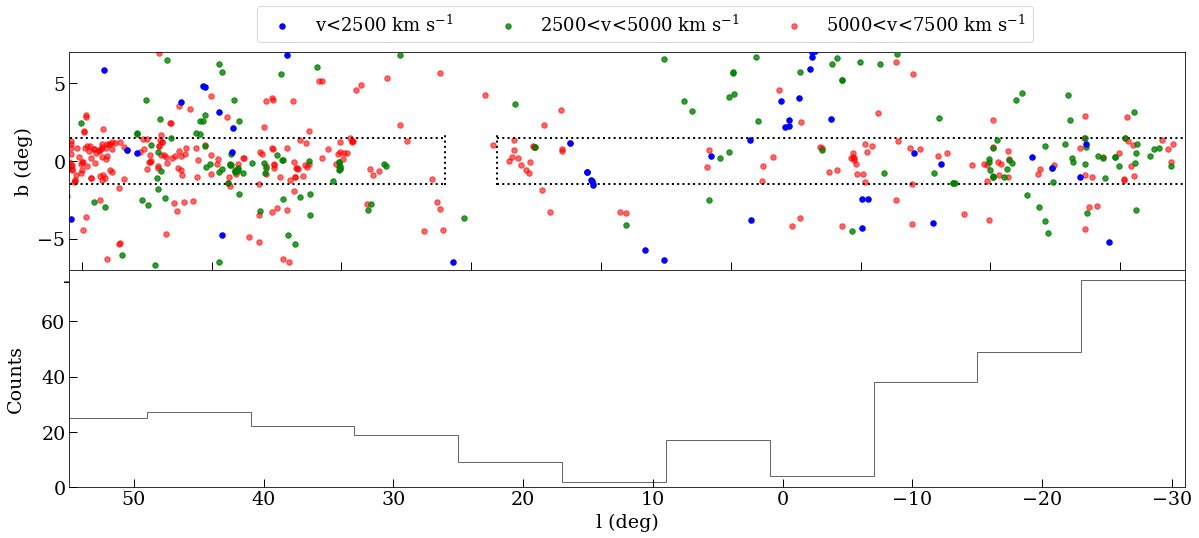}
    \caption{The distribution of galaxies in the Galactic coordinates. The region explored with LV--SMGPS survey is marked by the black dotted rectangles. The circles inside the rectangles ($-1.5 < b < 1.5$) represent the detections from the LV--SMGPS survey while the circles outside the rectangle ($b < -1.5; b> 1.5 $) represent HIZOA only detections. The Galaxies are colour coded by velocity range, as follows: the 300 -- 2500 \kms \ velocity range is shown in blue, 2500 -- 5000 \kms \ in green, and 5000 -- 7500 \kms \ in red.}
    \label{fig:l_b}
\end{figure*}

\begin{figure}
    \centering
    \includegraphics[width=0.95\linewidth]{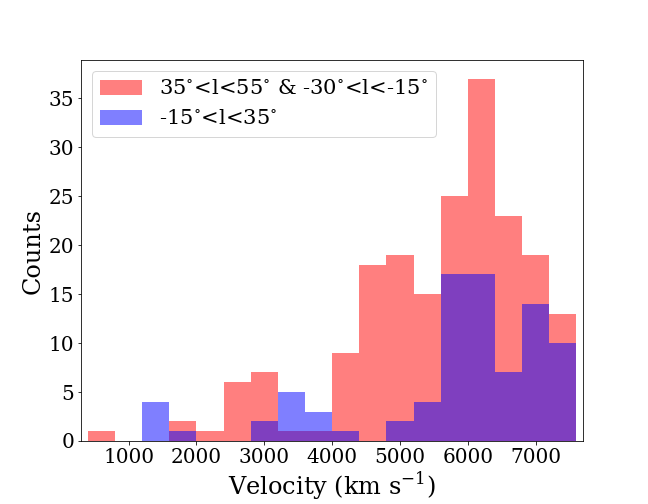}
    \caption{Histogram of velocities. The blue histograms represent the galaxies in the longitude range, $345^{\circ} <\ell< 35^{\circ}$ (LV) and the red histograms represent the longitude range, $329^{\circ} <\ell< 345^{\circ}$ and $35^{\circ} <\ell< 55^{\circ}$}
    \label{fig:vel_hist}
\end{figure}

\begin{figure}
    \centering
    \includegraphics[width=1.03\linewidth]{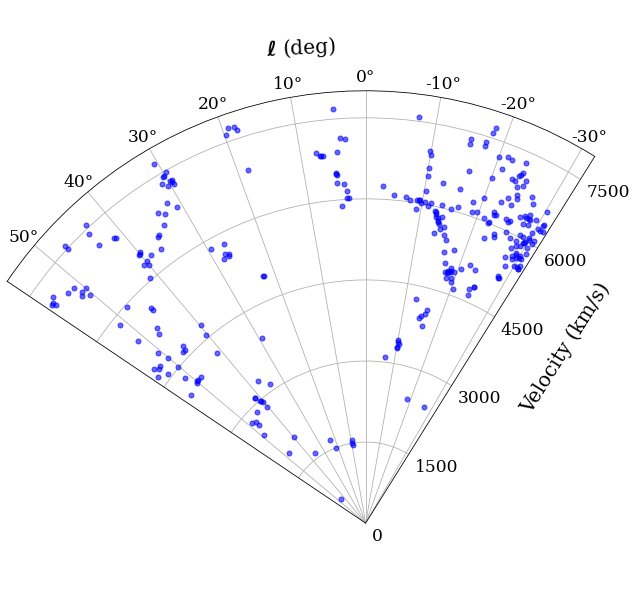}
    \caption{ Redshift wedge of $\Delta b = 3^{\circ}$ along the Galactic longitude for the LV--SMGPS detections out to \vhel\ $<$ 7500 \kms.}
    \label{fig:Wedge plot}
\end{figure}

\section{Results}

\label{sec:results}

\subsection{Large-scale structure}
In this section, we investigate the large scale structures in and around the Local Void as uncovered by the LV--SMGPS survey. Figure \ref{fig:l_b} (upper panel) displays the spatial distribution of \HI detections in Galactic coordinates for  \vhel $ <$ 7500 \kms. The region explored with LV--SMGPS survey ($|b| <$ 1.5$^{\circ}$) is marked by the black dotted rectangle and the circles inside this region represent SMGPS detections. The galaxies are colour coded by \vhel \ range, with blue, green, and red for 300 -- 2500 \kms, 2500 -- 5000 \kms, and 5000 -- 7500 \kms respectively. The lower panel of Fig.  \ref{fig:l_b} shows the number of galaxies as a function of Galactic longitude, each bin consisting of $\Delta \ell = 8^{\circ}$. It is quite obvious that the galaxy density is considerably lower in the longitude range, 345$^{\circ}$ $< \ell <$ 35$^{\circ}$ compared to other regions, especially for galaxies with \vhel\ below 5000 \kms \ -- the region corresponding to the Local Void. These findings are consistent with \citet{kraan08}, where they use HIZOA survey and find that LV consists of a huge under-dense region out to \vhel\  $\sim 6000 \ \kms$ from about $345^{\circ}$ to $45^{\circ}$ in longitude and $-30^{\circ}$ to about $+45^{\circ}$ in latitude. Despite the significantly enhanced sensitivity and resolution of the LV--SMGPS survey in comparison to HIZOA, which enables the detection of low mass dwarfs (up to 10$^{8}$ M$_{\odot}$ at 7500 \kms), we still find the LV to be under-dense. Nevertheless, the Void is not completely devoid of galaxies. 

The number of galaxies in the region 329$^{\circ}$ $< \ell <$ 345$^{\circ}$ is significantly higher compared to the other edge of the Void ($\ell > 35^{\circ}$).  In particular, there is a high density of galaxies with \vhel\ ranging from 5000 \kms\ to 7500 \kms, possibly connecting with a part of the GA wall, also known as the Norma Wall or the Norma supercluster \citep[][Steyn et al. 2023]{woudt01, woudt04, Radburn06, staveley16}. 

In Fig. \ref{fig:vel_hist}, we display the velocity histogram of the detections. The longitude range $345^{\circ} < \ell <35^{\circ}$ dominated by void galaxies is shown by blue histograms since we find a lower galaxy density (a factor of 3.5 for galaxies with $\vhel < \ 6000 \ \kms $ ) in this longitude range and the redshift distribution  of galaxies from other regions ($329^{\circ} < \ell <345^{\circ}$ and $35^{\circ} <\ell <55^{\circ}$) is shown by red histograms. The velocity histograms (blue) of the LV show that the border of the Void is approximately around 6000 \kms. 

Further, there seem to be peaks around $\sim$ 1500 km~s$^{-1}$ and $\sim$ 3500 km~s$^{-1}$ suggesting possible substructures in the Void. The velocity histograms of other regions (red)  do not have structure, except for a peak at 5500 -- 6500 \kms\, which  is related to the GA overdensities.

\begin{figure*}
    \centering
    \includegraphics[width=0.95\linewidth]{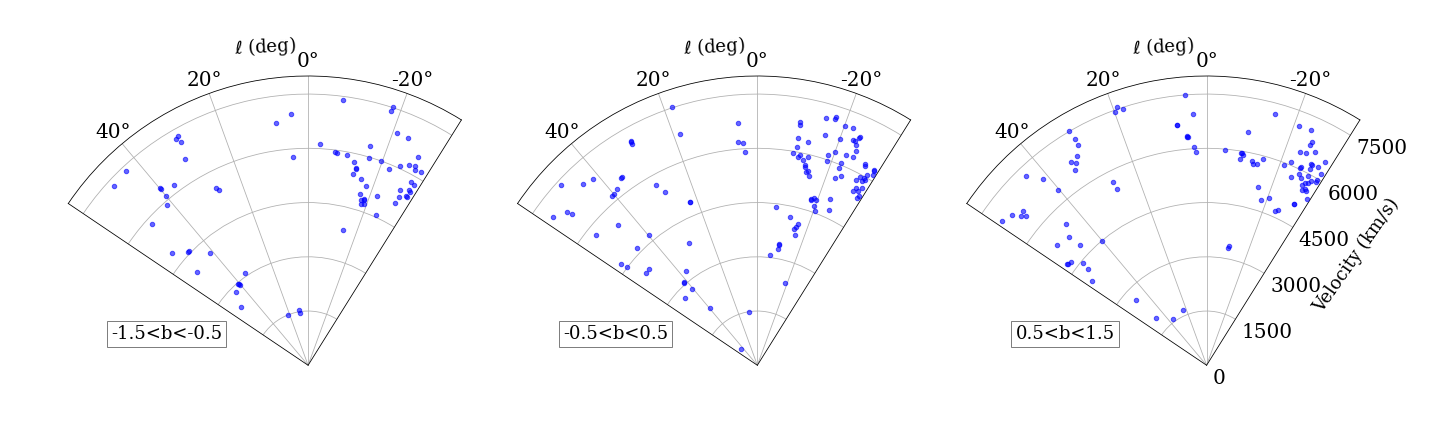}
    \caption{Redshift wedge plots for the \HI detected galaxies in the LV--SMGPS survey in the latitude ranges -1.5$^{\circ}$ $b<$ -0.5$^{\circ}$, -0.5$^{\circ}$ $b<$ 0.5$^{\circ}$, and 0.5$^{\circ}$ $b<$ 1.5$^{\circ}$ respectively.}
    \label{fig:Wedge_b}
\end{figure*}

\subsection{The Local Void and its borders}
\begin{figure}
    \centering
    \includegraphics[width=0.85\linewidth]{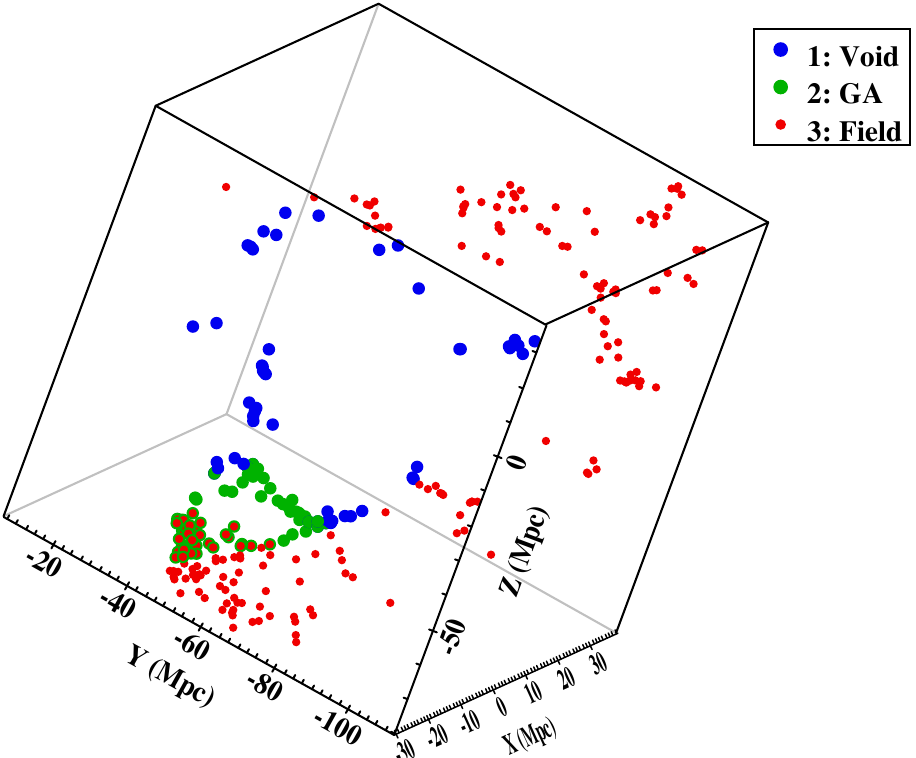}
    \caption{3D distribution of galaxies in Cartesian coordinates. The blue circles represent the galaxies presumably residing in the LV (and it's border), the red circles are galaxies beyond the Void, and the green circles are the galaxies that lie at the edge of Great Attractor wall.}
    \label{fig:3D_cartesian}
\end{figure}

 Figure \ref{fig:Wedge plot} displays the distribution of the detected galaxies in a redshift slice along Galactic longitudes for the width of the SMGPS, $|b|<$ 1.5$^{\circ}$, out to the investigated volume of  $\vhel < 7500 \ \kms$. The distribution of H{\sc i} detections is far from uniform, revealing various distinct features. The most prominent clearly is the LV, which at these low latitudes shows severe deficiency of galaxies in the region $-15^{\circ} < \ell <35^{\circ}$ with $\vhel <$ 6000 \kms.
 %Some signs of structure within the Local Void are visible, reminiscent of the tendrils in the wider HIZOA Galactic Bulge survey \citep[Kraan-Korteweg et al. 2022 in prep]{kraan08}. 
 %The detection rate goes up significantly in the region  -30$^{\circ}$ $< \ell <$ -20$^{\circ}$ and 5500 \kms \  $<$ v $<$7000  \kms \ which is possibly connecting with part of the Great Attractor wall ( also referred to as the Norma Wall or the Norma supercluster; \citep{kraan96, woudt04, woudt08, kraan11}). 
 In Fig. \ref{fig:Wedge_b}, the left, centre, and right panels show the distribution of the detected galaxies in a redshift slice in the latitude ranges, $0.5^{\circ} <b< 1.5^{\circ}$, $-0.5^{\circ} <b< 0.5^{\circ}$, and $-1.5^{\circ} <b< 1.5^{\circ}$ respectively. This demonstrates that the large scale structure is similar across the full width of the LV--SMGPS survey. 

%To further assess the significance of under-density of the Local Void, 

We now focus on the study the properties of galaxies as a function of local density, specifically categorizing galaxies based on their position i.e. whether they reside inside the Void, its border, or beyond the Void (hereafter referred to as the `field'). Hence, to classify LV-SMGPS detections into these three categories, we assess the boundary of LV. 
%Based on these results, we classify the detected galaxies into two categories, i.e. (i) galaxies in low density regions and (ii) galaxies in average density regions in the following method. 
We consider all galaxies with velocities higher than 6000 \kms \ to be in average density environment since the Local Void border is approximately at around 6000 \kms \ (from Fig. \ref{fig:vel_hist}). In addition, we consider all the galaxies in the longitude range, 40$^{\circ} < \ell <55^{\circ}$ to be of average density  with the regular distribution of galaxies from 300 -- 7500 \kms as seen in  Fig. \ref{fig:Wedge plot}. We convert the Galactic coordinates and the recession velocities into 3-D Cartesian coordinates to measure the actual distance between galaxies and allows us to classify Void galaxies, border galaxies, and  field galaxies more quantitatively. Fig. \ref{fig:3D_cartesian} shows the 3D distribution of  detected galaxies in Cartesian coordinates. The galaxy density in the central regions of the cube is very low, confirming the Local Void to be underdense.   The blue circles represent the galaxies presumably within the Void, the red circles represent the galaxies residing in average density regions, and the galaxies that are at the edge of the GA wall are shown by green circles.

%We also remove the galaxies that are at the edge of the Great Attractor (shown by green circles in Fig. \ref{fig:3D_cartesian}) from the sample of Void galaxies. This gives us a total of 55 galaxies in the Void.

\begin{figure}
    \centering
    \includegraphics[width=0.80\linewidth]{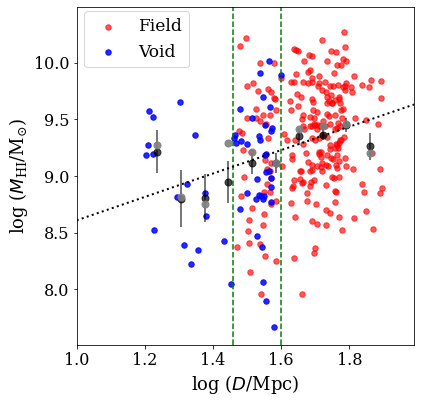}
    \caption{ Log-log plot of \HI mass versus the LV--centric distance. The Void and field galaxies are shown by blue and red circles. The green dashed lines represent the inner and outer border of Local Void. The black dotted line shows the linear regression over all galaxies. It has a slope of $\sim$ 1.03 $\pm$ 0.21.  The black and grey circles represent the mean and median \HI mass values per $log D$ bin.}
    \label{fig:HI_mass_dist}
\end{figure}
\begin{figure}
    \centering
    \includegraphics[width=0.85\linewidth]{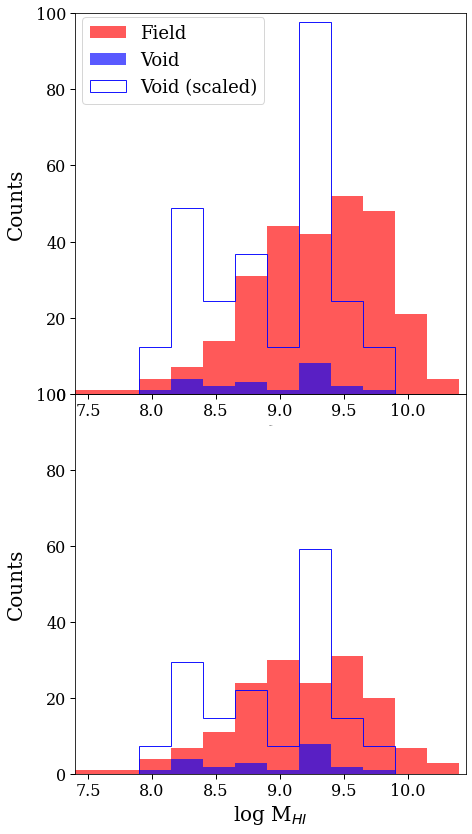}
    \caption{ \HI mass histograms for LV--SMGPS detections. The blue histograms represent the Void galaxies, red histograms the galaxies beyond the Void. Top panel: $ \vhel  < 7600 \ \kms$. Bottom panel: $\vhel  < 6000 \ \kms$.}
    \label{fig:HI_mass}
\end{figure}

\subsubsection{Extent of the LV at lowest latitudes}
\label{sec:LV_categories}
To determine the extent of the Void at the lowest Galactic latitudes, we estimate the centre of the Void as outlined by LV--SMGPS detections by assuming that it can be approximated reasonably well by the shape of a sphere. 
 We calculate the centre of sphere by taking the centroid of centres of the opposite points of sphere lying on great circles within Fig. \ref{fig:3D_cartesian}. We compute the distance of each galaxy from the fiducial Void centre and check whether a sudden increase in galaxy count occurs beyond a specific radius. Additionally, we examine whether there is a trend between \HI mass and the LV--centric radius.
 
 In Fig. \ref{fig:HI_mass_dist}, we display the resulting log-log plot of \HI mass (log  $M_{\rm HI}$) versus the distance to the Void centre (log  $D$). The blue and red circles represent the galaxies presumably lying in Void and field respectively.  The green dashed lines represent the inner (at log  $D \ = 1.46$ Mpc) and outer (log  $D= 1.60$ Mpc) boundaries of the LV. The density of galaxies inside the inner boundary is significantly lower compared to the density of galaxies outside this radius. This implies that the Void has a diameter of $\sim$ 58 Mpc. 
 This is consistent with the estimated dimensions of the LV previously derived from cosmicflows-3, measuring roughly 69, 51, and 60 Mpc \citep{tully19}. The cosmicflows-3 study uses distances and inferred peculiar velocities of nearby galaxies to reconstruct the distribution of overdensities and underdensities in the nearby universe.  We note that the LV doesn't have a simple shape \citep{rizzi17, tully19}
 
 We categorize galaxies according to their location with respect to the LV to study properties of galaxies as function of environment. There are 17 galaxies deep in the Void, i.e. inside the inner boundary (log $D \ < \ 1.46$ Mpc), 96 galaxies are at the border of the Void, i.e. between inner and outer boundary of the Void  (1.46 Mpc $<$ log $D <$ 1.60 Mpc), and the remaining 178 galaxies are in field, i.e. beyond the Void  (log $D \ > \ 1.60$ Mpc).

\subsection{HI mass variation with environment} 
In this section, we investigate the \HI mass as a function of LV--centric radius. In Fig. \ref{fig:HI_mass_dist}, we display log  $M_{\rm HI}$  versus the log  $D$. The black dotted line shows the linear regression of all galaxies with a slope of $\sim 1.03 \pm 0.21$. The linear regression has a large scatter $\sim 0.4$. The black and grey circles  represent the mean and median \HI mass values. The distribution of the points in Fig. \ref{fig:HI_mass_dist} reveals tendency of the \HI mass with LV--centric radius. The trend is also clear in the points of the outer envelopes of the distribution along log $D$. To assess the validity of this trend, we use the Spearman rank correlation coefficient and  Pearson correlation coefficient to analyze the relation between the \HI mass and the LV--centric radius. The Pearson correlation coefficient measures the strength of the linear monotonic correlation between two sets of data while the Spearman's correlation estimates the strength of monotonic relationships, regardless of whether they are linear or not. We calculated the Spearman's correlation coefficient and Pearson correlation coefficient and we find them to be very similar. We find that the Spearman correlation coefficient is 0.255 (p value $\sim$ 1.3 e$^{-5}$) and the Pearson correlation coefficient to be 0.251 (p value $\sim$ 1.4 e$^{-5}$), which means that the \HI mass and LV--centric radius are weakly correlated, but with a high significance. We divided the data into several sections with bin size 0.08 and the average/median value of each bin is marked in grey/black circles in Fig. \ref{fig:HI_mass_dist}. The binned Spearman and Pearson correlation coefficients for the mean values are 0.90 (p value $\sim$ 9 e$^{-4}$) and 0.91 (p value $\sim$6 e$^{-4}$) and the binned Spearman and Pearson correlation coefficients for the median values are 0.63 (p value $\sim$ 0.06) and 0.73 (p value $\sim$ 0.026). The correlation appears to be much stronger in the binned data sets.

As a further assessment of the \HI mass dependency of local density, i.e. environment, we consider the \HI mass distribution of Void galaxies with log  ($D$/Mpc) $<$ 1.46, and galaxies at the border of Void and beyond Void with log  ($D$/Mpc) $>$ 1.46. We test whether there are differences in \HI mass distributions within the Void compared to those beyond the Void. The top panel in Fig. \ref{fig:HI_mass} shows the \HI mass histogram of all detected galaxies. Void galaxies are represented by blue and galaxies beyond the Void are represented by red histograms. The average \HI mass of Void galaxies is $\sim$ 10$^{8.93}$ M$_{\odot}$ while the average \HI mass of galaxies in average density regions is $\sim$ 10$^{9.3}$ M$_{\odot}$, corresponding to a difference of $\sim$ 0.4 dex. We perform a Kolmogorov-Smirnov (KS) test which gives a probability of 0.015 for the galaxies in this Void and beyond the Void being drawn from the same distribution. The test implies a difference at the 2.5 $\sigma$ level, which is not statistically significant. However, the samples probe different volumes: LV $<$ 6000 $\kms$, while the whole sample is delimited by 7500 \kms. We reinvestigate by adopting the same volume limit. Therefore, we remove all the galaxies with velocities higher than $6000 \ \kms$ from the sample of field galaxies to keep same redshift range for Void galaxies and galaxies beyond the Void to compare their \HI mass distributions. In the bottom panel of Fig. \ref{fig:HI_mass}, we show the \HI mass of detected galaxies with \vhel  $< 6000 \ \kms$. The average \HI mass of galaxies beyond the Void is $\sim$ 10$^{9.18}$ M$_{\odot}$ corresponding to a difference of $\sim$ 0.25 dex between Void galaxies and field. The KS test gives a p value of $\sim$ 0.14 for both the galaxies in the Void and beyond Void to be drawn from the same distribution. This does not support the hypothesis that the two samples are being drawn from different populations. 

However, it should be highlighted that the Void doesn't host any massive galaxies, i.e. galaxies with M$_{\rm HI} >$ 10$^{9.7}$ M$_{\odot}$. Moreover, the number of galaxies drops significantly for galaxies with \HI mass higher than 10$^{9.4}$ M$_{\odot}$. This is consistent with results from studies that reconstruct the cosmic web using spectroscopic or photometric redshifts, which show that galaxies with high stellar mass are more likely to be located closer to filaments and walls than their less massive counterparts in low-density regions \citep{Chen17, Kraljic18, Laigle18, Luber19}. The observed correlation between H{\sc i} mass and stellar mass \citep{Parkash18, Catinella18} further points to an expectation of fewer galaxies with higher H{\sc i} masses in voids, in agreement with our results. Furthermore, our results are also consistent with the numerical simulations that predict that the mass function in voids is steeper than in walls and filaments, i.e. the number of massive haloes is depressed compared to that of the low mass haloes \citep{gottlober03, alfaro20, habouzit20, rosas22}.

\section{clustering of H{\sc i} selected void galaxies}
\label{sec:clustering}
The galaxy clustering was found to be a strong function of environment \citep[e.g.][]{abbas07, pujol17, paranjape18}. In the context of hierarchical cosmological structure formation scenario, voids are expected to have a substructure consisting  of sub voids, tendrils, filaments, etc \citep{vande93, sheth04, vande11, aragon13, Odekon18}. Hence, void galaxies are not necessarily isolated at small scales ($ \lesssim $ 1 Mpc) but their density of galaxies is very low on the scale of voids. The small scale clustering of galaxies in voids was not found to differ from clustering of galaxies in average density environments. For example, \citet{szomoru96} found that the clustering of H{\sc i} detected galaxies within 1 Mpc of galaxies in the void is similar to clustering in average density regions (within a factor of two). \citet{abbas07} study environmental dependence of clustering for  galaxies in the Sloan Digital Sky Survey (SDSS) and they find that the relation between clustering strength and density is not monotonic. They find that galaxies in the void centers are more clustered than those in the void outskirts,  but the latter are less clustered than those at moderate overdensities. 

\citet{pustilnik11} and \citet{pustilnik19} studied clustering of optically selected void galaxies and they find that around 17.5 percent of the void galaxies form pairs, triplets, quadruplets or larger groups.  \citet{kreckel12} observed 55 galaxies ($-16^{m}.1 \ > M_{r} > \ -20^{m}.4$) in voids using the Westerbork Synthesis Radio Telescope (WSRT). They detect \HI in 41 of the galaxies and they identify 18 \HI rich companions  in pairs\slash triplets, a total of 34 out of 59 $\sim$ 57 $\%$, and they suggest that the small-scale clustering of galaxies in voids is very similar to that in higher density regions. They have used sky separations of 600 kpc and velocity differences of 200 \kms \ due to their observational constraints. \citet{kurapati20} obtained \HI observations of 24 dwarf galaxies ($M_{B} > -16$ mag) in the Lynx-Cancer void and they find that 13 galaxies of 31 (including previous detections) i.e. 42 $\%$ were part of pairs or triplets.

\subsection{Clustering in LV and its surroundings}
We use the LV sample to study the clustering properties of \HI selected galaxies. We compare the small-scale clustering for galaxies residing in different local densities: the least dense regions (deep in LV), moderately underdense regions (border of the LV), average density regions (field excluding GA wall), and moderately overdense regions (GA wall).  Galaxies within the region of the GA wall (as seen in Fig. \ref{fig:3D_cartesian}) are categorized as being in a moderately overdense region, as the density of galaxies is clearly higher compared to regions with average density (see Fig. \ref{fig:Wedge plot}). For the remaining galaxies, we have used the categorization approach described in section \S \ref{sec:LV_categories}, based on the LV-centric distance, which divides galaxies into those deeply embedded within the Void (log $D < 1.46$), those on the border of the Void ($1.46 < $log $D \ < 1.60$), and galaxies in the field (log $D \ > 1.60$).

 We use sky separations of 600 kpc and velocity differences of 200 \kms \ to study the small scale clustering of \HI selected void galaxies to enable a direct comparison with previous studies  \citep{kreckel12, kurapati20}. If we consider only the galaxies that are in deep in the LV (log  $D < 1.46$ in Fig. \ref{fig:HI_mass_dist}), there are a total of 17 galaxies and 8 galaxies ($\sim$ 47 $\%$) within 600 kpc and  200 \kms, with 6 of them having one neighbour, one with two neighbours,  six with three neighbours, and two with four neighbours within 600 kpc and 200 \kms \, which is  consistent with the previous studies \citep{kreckel12,kurapati20}. 

We find that 22 out of 60 galaxies ($\sim$ 36.6$\%$) at the border of the Void and 26 out of 134 galaxies ($\sim$ 19.4$\%$) in the field (excluding the GA wall) are found within 600 kpc and 200 km/s, both of which show lower clustering compared to the clustering in the Void center. On the other hand, for a total of 80 galaxies in the GA wall, 43 ($\sim$ 53.8 $\%$) were observed within  600 kpc and  200 \kms . This is higher than the small-scale clustering close to the Void center and its outskirts, as well as the field.  This indicates that the small-scale clustering of the least dense regions (i.e center of LV)  is higher than that of moderately underdense regions (LV border and field), but is lower than the small-scale clustering of the moderately overdense regions (GA wall). This is in agreement with \citet{abbas07}, who found that the galaxies in the least dense regions are more clustered than galaxies in moderately underdensities, but galaxies in moderate underdensities are less clustered than those at moderate overdensities.

  %They found that the least dense regions are more clustered than galaxies in moderate underdensities, but the latter are less clustered than those 

%To study small-scale clustering of Void galaxies, galaxies in average density regions (field without the GA wall), and galaxies in moderately overdense regions (GA wall), we use the following method to distinguish different environments: We use all the galaxies in GA wall region (from Fig. \ref{fig:3D_cartesian}) as the galaxies in moderately overdense regions.  This is mainly because the density of galaxies at the GA wall region is clearly much higher than galaxy density in average density regions (see Fig. \ref{fig:Wedge plot}). In the remaining galaxies, we use the categorization from \ref{fig:HI_mass_dist} as described in section {\bf 4.2.} based on log D, i.e. galaxies residing deep in the Void (log  $D < 1.46$) , the border of Void ($1.46 < $log  $D \ < 1.60$) , and the galaxies in field (log  $D \ > 1.60$.

%We estimate the small-scale clustering for galaxies residing in the LV, GA wall, and in the field excluding GA.
\newcolumntype{R}[1]{>{\RaggedLeft\arraybackslash}p{#1}}
\newcolumntype{L}[1]{>{\RaggedRight\arraybackslash}p{#1}}
\newcolumntype{C}[1]{>{\Centering\arraybackslash}p{#1}}

\begin{table*}
\caption{\HI parameters of galaxies in groups.}
\begin{tabular}{L{0.6cm}C{1.0cm}L{0.4cm}L{2.1cm}R{0.8cm}R{0.75cm}C{0.65cm}R{0.55cm}R{0.55cm}R{0.9cm}R{0.9cm}R{0.85cm}R{0.5cm}R{0.7cm}}
\hline
Group & Region & No & Name &  $\ell$ & $b$ & \vhel & $w_{\rm 20}$ & $w_{\rm 50}$ & log($M_{\rm HI}$) & log($M_{\rm dyn}$) & log($M_{\rm halo}$)&  S & $d_{\rm proj}$  \\
 &  & &  SMGPS-HI- & deg & deg & \kms & \kms & \kms & log(\msun) & log(\msun) & log(\msun) & & \kpc \\
 (1)& (2) & (3) & (4) & (5) & (6) & (7) & (8) & (9) & (10) & (11) & (12) & (13) & (14) \\
\hline
\multirow{3}{*}{F1} & \multirow{3}{*}{Field} & a &  J160121-532553 & 328.98 & -0.45 & 5477 & 245 & 194 & 9.44 & 10.49 & \multirow{3}{*}{11.34} & \multirow{3}{*}{0.028} & \multirow{3}{*}{577} \\
& & b & J155952-530455 & 329.04 & -0.04 & 5545 & 340 & 281 & 9.83 & 11.20 & & & \\
& & c & J160114-531146 & 329.12 & -0.26 & 5487 & 129 & 113 & 8.91 & 10.48 & & & \\
\hline
\multirow{3}{*}{F2} & \multirow{3}{*}{Field}& a & J161709-521624 & 331.52 & -1.21 & 5162 & 207 & 167 & 9.58 & 10.48 & \multirow{3}{*}{10.70} & \multirow{3}{*}{0.037} 
 & \multirow{3}{*}{371}\\
& & b & J161621-520100 & 331.61 & -0.94 & 5160 & 87 & 72 & 8.76 & 9.75 & & &  \\
& & c & J161738-520111 & 331.75 & -1.08 & 5189 & 127 & 95 & 9.15 & 10.17 & & &  \\
\hline
\multirow{6}{*}{F3} & \multirow{6}{*}{Field} & a & J160803-505214 & 331.45 & 0.78  & 6290 & 136 & 114 & 9.27 & 10.18  & \multirow{6}{*}{12.01} & \multirow{6}{*}{0.028} & \multirow{6}{*}{1399} \\
& & b & J160728-503434 & 331.58 & 1.06 & 6385 & 89 & 74 & 8.99 &  9.63 & & &  \\
& & c & J160942-503349 & 331.85 & 0.83 & 6391 & 179 & 164 & 9.64 & 10.57 & & &  \\
& & d & J160914-501912 & 331.96 & 1.06 & 6456 & 164 & 127 & 9.91 & 11.24 & & &  \\
& & e & J161007-501128 & 332.15 & 1.06 & 6251 & 223 & 185 & 10.01 & 11.79 & & &  \\
& & f & J161210-501742 & 332.32 & 0.76 & 6384 & 267 & 232 & 9.71 & 11.25 & & &  \\
\hline
\multirow{7}{*}{B1} & \multirow{7}{*}{Border} & a & J165630-445219 & 341.36 & -1.06 & 4895 & 85 & 70 & 9.23  & 9.57 & \multirow{7}{*}{11.86} & \multirow{7}{*}{0.036} & \multirow{7}{*}{1407}\\
& & b & J165259-441844 & 341.40 & -0.22 & 4929 & 138 & 138 & 8.99 & 10.12 & & &  \\
& & c &  J165323-441909 & 341.44 & -0.28 & 4973 & 314 & 287 & 10.22 & 11.56 & & &  \\
& & d & J165549-441534 & 341.76 & -0.58 & 4896 & 134 & 71 & 8.82 & 9.58 & & &  \\
& & e & J165633-442106 & 341.77 & -0.74 & 4776 & 176 & 162 & 9.36 & 10.36 & & &  \\
& & f & J165639-440511 & 341.99 & -0.59 & 4903 & 356 & 324 & 10.15 & 11.47 & & &  \\
& & g & J165608-440041 & 341.99 & -0.47 & 4870 & 184 & 165 & 9.40 & 10.47 & & &  \\
& & h & J165852-435329 & 342.39 & -0.78 & 4868 & 46 & 46 & 8.37 & 9.41 & & &  \\
\hline
\multirow{3}{*}{B2} & \multirow{3}{*}{Border} & a & J171053-402428 & 346.51 & -0.44 & 5730 & 65 & 45 & 8.66 & 9.22 & \multirow{3}{*}{9.81} & \multirow{3}{*}{0.016} & \multirow{3}{*}{288} \\
& & b & J171120-402132 & 346.60 & -0.48 & 5755 & 93 & 72 & 8.29 & 9.11  & & & \\
& & c & J171150-401947 & 346.68 & -0.54 & 5795 & 93 & 58 & 8.93 & 9.54 & & &  \\
\hline
\multirow{5}{*}{V1} & \multirow{5}{*}{Void} & a & J172253-383451 & 349.34 & -1.27 & 3380 & 91 & 77 & 8.52 & 9.46 & \multirow{5}{*}{11.06} & \multirow{5}{*}{0.013} & \multirow{5}{*}{1100}\\
& & b & J171906-373147 & 349.78 & -0.06 & 3411 & 126 & 82 & 9.57 & 10.11 & & &  \\
& & c & J172058-374204 & 349.85 & -0.46 & 3448 & 163 & 124 & 9.27 & 10.47  & & &  \\
& & d & J172109-374256 & 349.86 & -0.50 & 3319 & 219 & 191 & 9.52 & 10.63 & & &  \\
& & e & J172031-373608 & 349.88 & -0.33 & 3298 & 127 & 112 & 9.19 & 10.44 & & &  \\
\hline
\multirow{4}{*}{B3} & \multirow{4}{*}{Border} & a & J175258-242216	& 4.76 & 0.94 & 6459 & 266 & 224 & 10.21 & 11.61 & \multirow{4}{*}{11.71} & \multirow{4}{*}{0.54} &  \multirow{4}{*}{225}\\
& & b & J175320-241850	& 4.85 & 0.90 & 6477 & 186 & 163 & 9.68 & 10.66  & & & \\
& & c & J175330-241811	& 4.88 & 0.87 & 6325 & 186 & 139 & 9.46 & 10.41  & & & \\
& & d & J175325-241645	& 4.89 & 0.90 & 6508 & 163 & 139 & 9.58 & 10.47  & & & \\
\hline
\multirow{3}{*}{V2} & \multirow{3}{*}{Void} & a & J180808-213547 & 8.90 & -0.71 & 1466 & 88 & 69 & 8.37 &  9.53  & \multirow{3}{*}{11.17} & \multirow{3}{*}{0.02} & \multirow{3}{*}{320} \\
&  & b & J181051-213408 & 9.23 & -1.25 & 1564 & 337 & 288 & 8.80 & 11.12  & & & \\
&  & c & J181149-213318 & 9.35 & -1.44 & 1511 & 183 & 142 & 8.83 & 10.07 & & & \\
\hline
\multirow{3}{*}{F4} & \multirow{3}{*}{Field} & a & J192836+195742 & 54.84 & 1.18 & 7076 & 241 & 191 & 9.84 & 10.91 & \multirow{3}{*}{10.98} & \multirow{3}{*}{0.008} & \multirow{3}{*}{923}\\
& & b & J192745+200537 & 54.86 & 1.42 & 7005 & 135 & 99 & 9.06 & 9.90 & & &  \\
& & c & J192956+200923 & 55.16 & 1.00 & 7079 & 93 & 83 & 8.89 &  9.68 & & & \\
\hline
\label{table:groups}
\end{tabular}
\end{table*}

\begin{figure}
    \centering
    \includegraphics[width=1.03\linewidth]{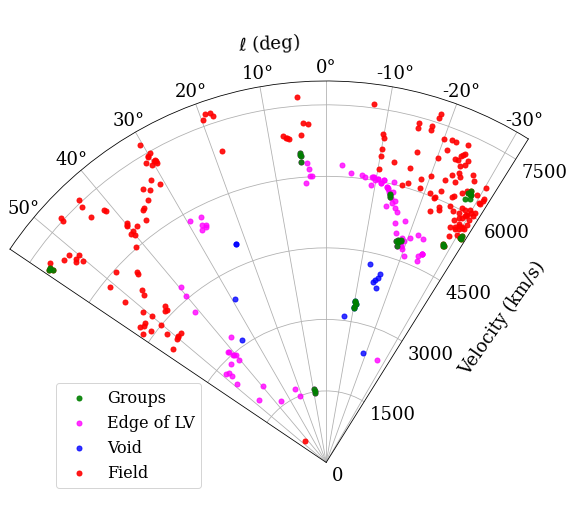}
    \caption{Redshift wedge for the LV-SMGPS detections. The blue circles represent the galaxies residing deep  within the LV (log  $D < 1.46$), magenta galaxies at the border of LV ($1.46 <$ log  $D < 1.60$), and red field galaxies. The galaxies residing in groups of galaxies are marked in green.}
    \label{fig:wedge_group}
\end{figure}

\begin{figure*}
    \centering
    \subfloat[Void group V1 (5 galaxies).]{\includegraphics[width=0.97\textwidth]{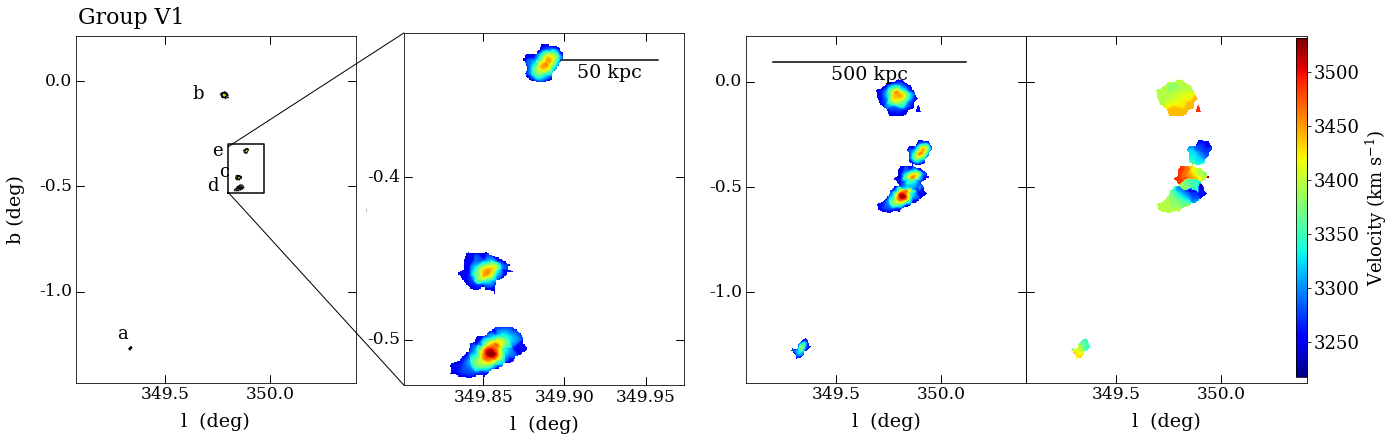}\label{fig:V1}}\\
    %\vspace{\baselineskip}
    \subfloat[Void group V2 (3 galaxies).]{\includegraphics[width=0.44\textwidth]{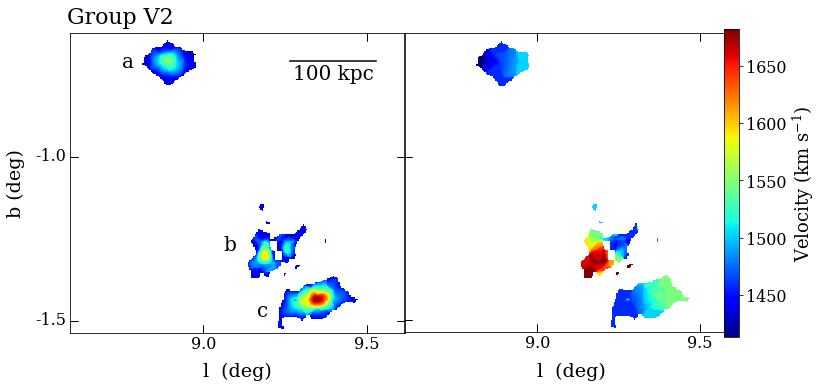}\label{fig:V2}}
    \subfloat[Border group B2 (3 galaxies).]{\includegraphics[width=0.49\textwidth]{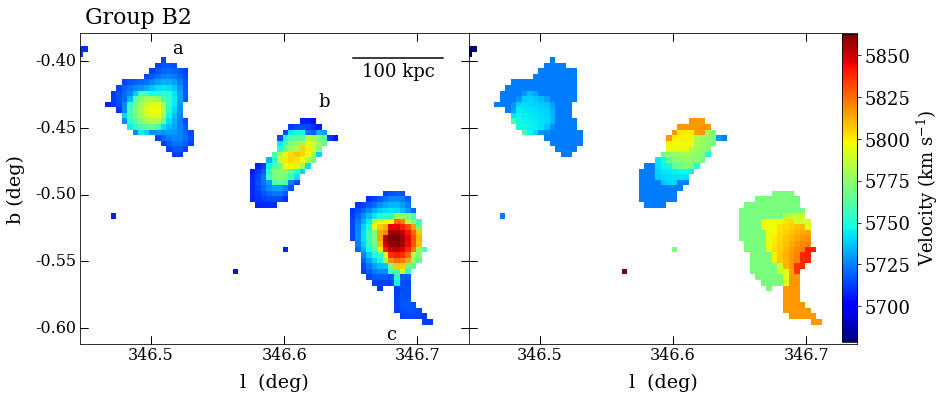}\label{fig:B2}}
    \vspace{\baselineskip}
    \subfloat[Border group B1 (8 galaxies).]{\includegraphics[width=0.97\textwidth]{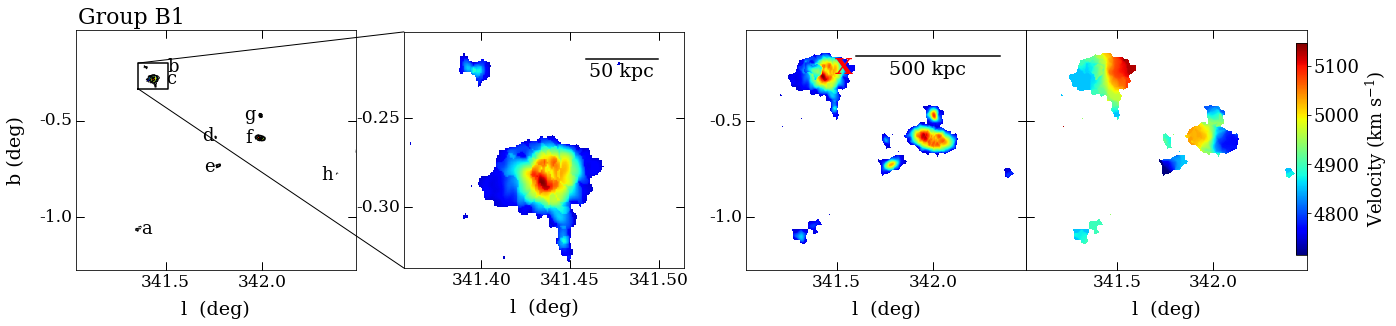}\label{fig:B1}}
    \vspace{\baselineskip}
    \subfloat[Border group B3 (4 galaxies).]{\includegraphics[width=0.59\textwidth]{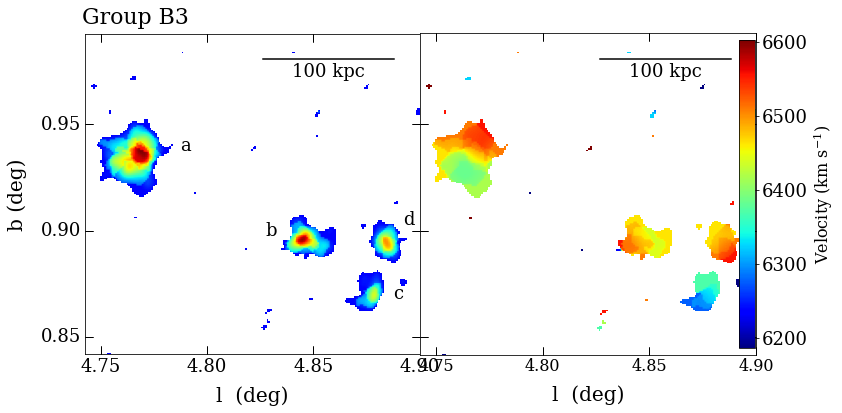}\label{fig:B3}}
    \caption{Galaxy groups in Void and border of Void.}
    \label{fig:Void_Border_groups}
\end{figure*}
\begin{figure*}
    \centering
    \includegraphics[width=0.49\linewidth]{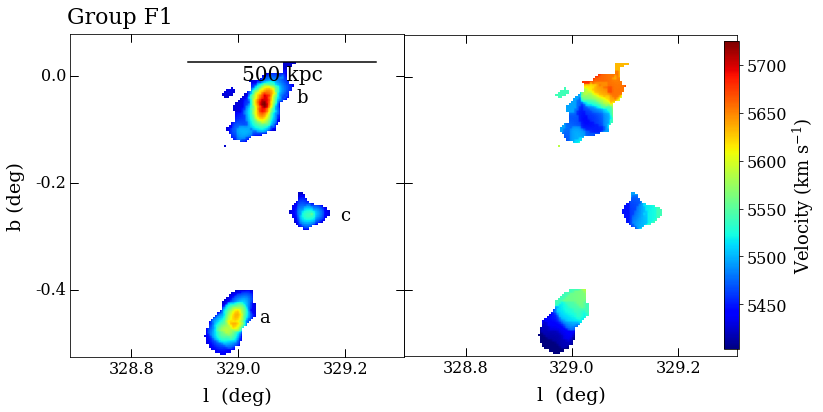}
    \includegraphics[width=0.49\linewidth]{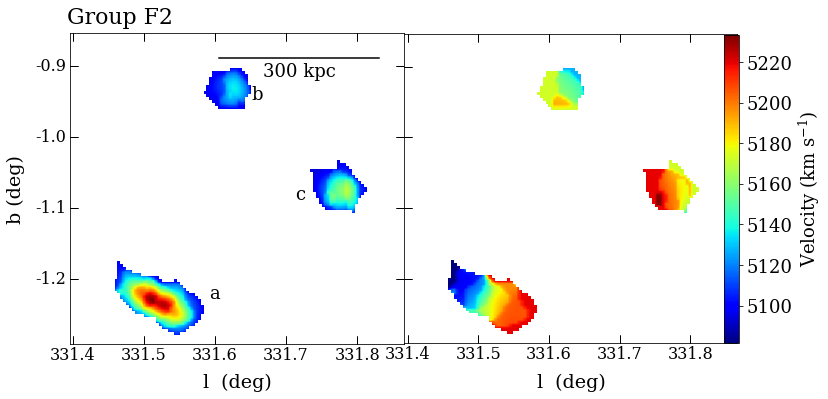} \\
    \includegraphics[width=0.49\linewidth]{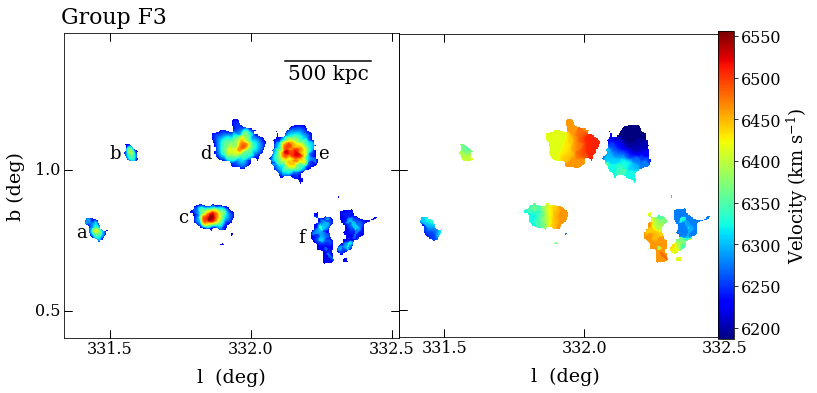}
    \includegraphics[width=0.49\linewidth]{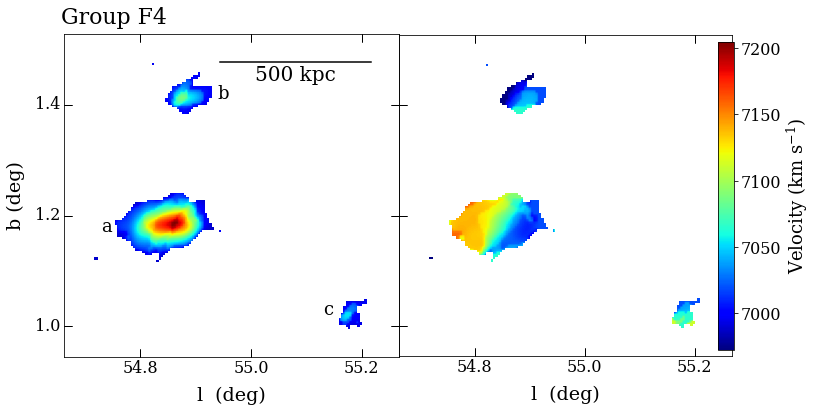} 
    \caption{ Upper panel: Loose group of 3 galaxies in average density environment (towards the edge of Great Attractor). Lower panel: Loose group of 3 galaxies in average density environment. The galaxies are located at the proper position, but are magnified by a factor of five to reveal the details of \HI distribution.}
    \label{fig:Field_groups}
\end{figure*}

\subsection{Potential galaxy groups}
Voids are expected to  exhibit intricate fine substructure consisting of sub-voids, tendrils, filaments. Numerical simulations predict that galaxies in voids preferentially lie within the filamentary void substructures. In fact, some earlier observational studies have found indications for such filamentary substructure in voids \citep[e.g.][]{zitrin08, beygu13}. For example, \citet{beygu13} reported three interacting galaxies embedded in a common \HI envelope, which they hypothesized to be an assembly of a filament in a void. \citet{chengalur17} find an approximately linearly aligned triplet of gas-rich dwarfs with large-scale velocity continuity, which they argue is consistent with the galaxies lying along a filament. This can be interpreted as a consequence of the structure formation proceeding more slowly. 

To explore if we find filamentary substructures in the LV, we search for galaxy group candidates in the LV--SMGPS data. We mainly use the projected separation between galaxies and the difference between recession velocities for the identification of groups. We find a total of nine possible groups, some of these are tight groups, while others may still be in the process of assembling. Figure \ref{fig:wedge_group} shows the position of the potential groups in a redshift wedge. The blue dots represent galaxies residing deep in the Void (log  $D < 1.46$) , magenta the galaxies at the border of Void ($1.46 < $ log  $D \ < 1.60$) while red indicates the galaxies in field (log  $D \ > 1.60$; see Fig. \ref{fig:HI_mass_dist}). Galaxies conforming potential groups are marked in green. Two of the groups (with 5 and 3 members) are located in the LV, three of them (with 8, 4, and 3 members respectively) lie at the border of the LV, and the other four (6 and 3 members) are found in the field. The velocities of the group members all lie within 200 \kms.

We estimate the dynamical mass of each group member following \citet{deblok14} using 
\begin{equation}
    M_{\rm dyn} = \frac{R}{G} \bigg(\frac{w_{50}}{2\sin i}\bigg)^{2} ,
\end{equation}
where $R $  and $i$ are radius of the \HI disc and inclination angle of galaxy. The inclination angle is determined by using $\cos^{2}(i) = (b^{2}-\theta_{b}^{2})/ (a^{2} - \theta_{a}^{2})$, where a and b are the major and minor axis of the galaxy, $\theta_{a}$ and $\theta_{b}$ are the sizes of the synthesis beam. 

To quantify the compactness of groups, we define the compactness parameter, $S$,  following \citet{duplancic17}, \begin{equation} 
S = \sum^{N}_{i=1}  \frac{R^{2}}{(d_{\rm proj}/2)^{2}},
\end{equation}
where $N$ is the number of galaxies in the group and  $d_{\rm proj}$ is the maximum distance between group members. \citet{duplancic13, duplancic17} have adopted a threshold $S > 0.03$ which corresponds to compactness of compact groups of galaxies. But, we note that they used optical diameters and here we are using \HI diameter.

%\begin{figure}
%\addtocounter{figure}{-1}
%    \centering
%    \includegraphics[width=1.05\linewidth]{B3_group4.png} 
%    \caption{ (e) Group of 4 galaxies at the edge of the Void. The galaxies are located at the proper position and are at the original size.}
%    \label{fig:B3}
%\end{figure}

Table \ref{table:groups} lists the \HI parameters of galaxies residing in groups. The columns are as follows:\\
{\it Column 1} - Name of group  \\
{\it Column 2} - Environment in which group was found  \\
{\it Column 3} - Group member identifier  \\
{\it Column 4} -  Name of galaxy, LV--SMGPS identifier giving RA and Dec \\
{\it Column 5 and 6 } -  Galactic longitude (deg) and latitude (deg)  \\
{\it Column 7} -   Optical heliocentric velocity, $V_{\rm hel}$ in \kms \\
{\it Column 8 and 9} -   Line widths $w_{\rm 50}$ and $w_{\rm 20}$ in \kms \\
{\it Column 10} - \HI mass in log M$_{\odot}$ \\
{\it Column 11} - Dynamical mass in log M$_{\odot}$ \\
{\it Column 12} - Total dynamical mass of the group in log M$_{\odot}$ \\
{\it Column 13} - Compactness parameter, S. \\
{\it Column 14} - Maximum projected distance between the members of group, $d{\rm proj}$ . \\
%\begin{figure*}
%    \centering
%    \includegraphics[width=0.97\linewidth]{V1_group5.png} \\
%    \includegraphics[width=0.97\linewidth]{B1_group8.png} \\
%    \includegraphics[width=0.46\linewidth]{V2_group3.png}
%    \includegraphics[width=0.51\linewidth]{B2_group3.png} \\
%    \caption{ Group of 5 galaxies inside the Void. First panel: The galaxies are located at the proper position and are at original size, The region outlined with a box is shown in the second panel. Third and fourth panels:  The \HI morphologies and velocity fields of galaxies in the group. The galaxies are located at the proper positions but are magnified by a factor of five to show the details of \HI distribution.}
%    \label{fig:T3_group5}
%\end{figure*}

%Cosmological simulations show that the voids are filled by low-density dark matter filaments creating substructures within their interior (van de Weygaert van Kampen 1993; Gottlo ̈ber et al. 2003; Colberg et al. 2005b; Springel et al. 2006) and that the galaxies in voids may preferentially lie within these filamentary void substructures \citep{sheth04, vande11, aragon13, rieder13}. This may indicate that the galaxies residing in voids are formed along these dark matter filaments, given that the simulations reveal that dark matter haloes are forming along them. 
\begin{figure}
    \centering
    \includegraphics[width=1.05\linewidth]{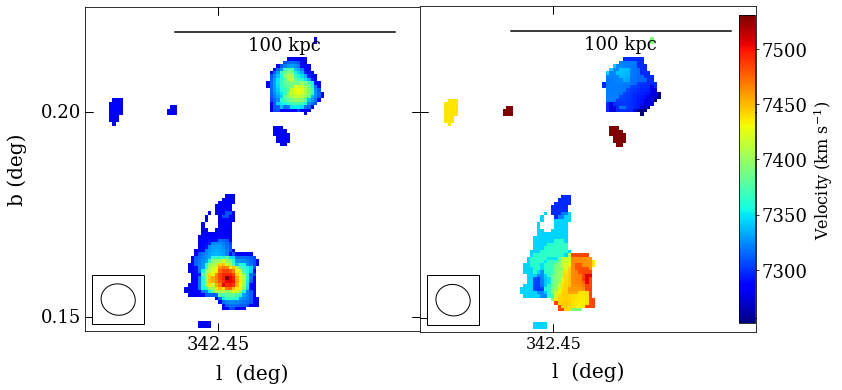}
    \caption{Pair of galaxies in average density environment.}
    \label{fig:pair}
\end{figure}

\subsubsection{Groups in the Void (V1--V2)}
Figure \ref{fig:Void_Border_groups} displays all the galaxy group candidates in the Void and at the border of Void. The top row (\ref{fig:V1}) shows the \HI intensity map of the group V1 which contains 5 galaxies. The left panel shows all 5 galaxies within one panel.  The maximum projected distance ($d_{\rm proj}$) between the group members is $\sim$ 1.1 Mpc. It has a fairly compact core (see box) with $d_{\rm proj} \sim$ 142.5 kpc. The next two panels show the HI morphologies and the velocity fields of the group galaxies. The galaxies are at their actual position but are magnified by a factor of five to visualize the details of their \HI distribution. The left panel of second row (\ref{fig:V2}) shows the HI intensity map and velocity field of a group of 3 galaxies in the Void. The \HI morphology of the galaxy J181051-213408 is peculiar, it is in the form of a \HI ring with a tail like feature in the north.  The maximum projected distance between the galaxies is 320 kpc. The approximately linear alignment of the galaxies in these groups (the compact core of V1 and the 3 galaxies of group V2 ) resembles galaxies arranged along intra-void filaments, which could signify  the ongoing growth of these galaxies along a filament inside the LV.

%\HI intensity map, However, the group of 3 galaxies outlined by a box (as shown in second panel) is more compact, with the maximum projected distance between them is 142.5 kpc. Third and fourth panels show the \HI morphologies and velocity fields of galaxies in the group. The galaxies are at the proper position but are magnified by a factor of five to show the details of \HI distribution. Figure \ref{fig:N3_group3} shows the \HI intensity maps and velocity fields of a group of 3 galaxies in the Void. The galaxies were magnified by  a factor of five to show the details of their \HI morphologies and velocity fields.  The approximately linear alignment of the galaxies in these groups (the compact group of 3 galaxies shown in second panel of Figure \ref{fig:T3_group5} and group of 3 galaxies shown in in Figure \ref{fig:N3_group3}) is consistent with galaxies lying along intra-void filaments, and this may suggest that we are witnessing the ongoing growth of these galaxies along a filament inside the Void.

\subsubsection{Groups at the edge of the Void (B1--B3)}
We identified three possible groups at the borders surrounding the Void. The group with the highest number of members, B1 with 8 galaxies is displayed in the third row (\ref{fig:B1}) of Fig. \ref{fig:Void_Border_groups}. The first panel is at fixed original scale. A zoom-in of the region outlined by a box is shown in the next panel. It is a very extended galaxy  (J165323-441909) of diameter $\sim$ 80 kpc and has a close companion (J165259-441844) at a projected distance of $\sim$ 35 kpc. The galaxy, J165323-441909 has a tail extending in the south direction, possibly due to gas accretion from a intra-void filament or due to a gas-rich minor merger. The third and fourth panel show the full group  with magnified \HI maps. The close companion galaxy, J165259-441844 is shown by `X'  as it is not visible in the magnified \HI map. The maximum projected distance between these galaxies is $\sim$ 1.4 Mpc within 200 \kms. However, the group of four galaxies (J165608-440041, J165639-440511, J165549-441534, and J165633-442106) within B1 is within $\sim$ 424 kpc. The close proximity in space and redshift may suggest that majority of these are group members, however, a few galaxies could be in the process of joining the system.

The right panel of second row (Fig. \ref{fig:B2}) shows group B2 of three galaxies at the edge of the Void, with $d_{\rm proj}$ $\sim$ 287 kpc.  Although not located within the Void but at its borders, the image also reveals a linear alignment of galaxies with large scale velocity continuity, consistent with galaxies lying along a filament.  

The bottom row (\ref{fig:B3}) shows their \HI distribution (morphology and velocity filed) of a compact group B3, which consists of four galaxies at the original scale. It  was originally detected as a single massive galaxy in the HIZOA survey (HIZOA J1753-24A). The maximum projected distance between the members of this group is around $\sim$ 225 kpc. The group of three galaxies (J175320-241850, J175325-241645, and J175330-241811)in the lower right is much more compact with the maximum projected distance of just $\sim$ 68 kpc. 

%Figure \ref{fig:B1} shows a group of 8 galaxies at the edge of Void. First panel shows galaxies at original scale. The region outlined by a box is shown in the second panel to show a large galaxy (J165323-441909) and its close companion (J165259-441844). The projected distance between these two galaxies is $\sim$ 35 kpc and the diameter of J165323-441909 is $\sim$ 80 kpc. The galaxy, J165323-441909 has a tail extending in the south direction, which could be due to gas accretion from a intra-void filament or due to a gas-rich minor merger. The third and fourth panel show the \HI distribution (magnified by a factor of five) and velocity fields of galaxies of this group of 8 galaxies. T

\subsubsection{Groups in field (F1--F4)}

The four candidates for groups of galaxies in average density regions are given in Fig. \ref{fig:Field_groups}. In all the panels, the galaxies are at their actual position but are magnified by a factor of five to visualize the details of their \HI distribution. Groups F1 (top left), F2 (top right), and F4 (bottom right) have 3 members each. Group F3 (bottom left) contains 6 galaxies, with $d_{\rm proj} \ \sim$ 1.4 Mpc. However, no further substructure was notable ($d_{proj} <$ 500 kpc).  %{\bf The group may still be in the process of assembly.} 
The \HI morphology of galaxy, J161210-501742 in this group is very interesting. It shows a \HI ring with $\sim$ 65 kpc diameter. The ring galaxies could have various origins such as collisional origin with an intruder galaxy \citep[e.g.][]{bait20}.
However, it is difficult to comment on the formation scenario of this ring galaxy as we do not have observations in optical/infrared wavelengths given its location in the ZoA and we did not find any nearby companions for this ring galaxy.  

Group F1 (top left)  and Group F2 (top rights) have three members and lie in the vicinity of the Great Attractor wall, with  $d_{proj} \sim$  578 and 371 kpc respectively. Group F4 also has three members with $d_{proj} \sim$ 923 kpc. These could be loose group of galaxies that may still be in the process of forming themselves. 

We also found a close pair of galaxies (Fig. \ref{fig:pair}) in the field with offsets of only 100 kpc and 100 \kms. We haven't found any compact groups (with $d_{proj} < $ 350 kpc) in average density environments in contrast to Void and the border of Void.

\section{Summary and Conclusions}
\label{summary}
We used the \HI  21 cm emission observations from the LV--SMGPS to study the \HI galaxy population in and around the LV along the inner Zone of Avoidance ($|b|< \pm \ 1.5^{\circ}$), including their \HI properties as a function of local environment. We detected and parameterised 291 galaxies and categorized the galaxies based on their environment. We find that 17 of the detected galaxies lie deep inside the LV, 96 galaxies are at the border of the LV, while the remaining 178 galaxies are in average density environments. The extent of the Void as traced by LV-SMGPS is $\sim$ 58 Mpc. It is severely under-dense over the longitude range 350$^{\circ}< \ \ell \ <$ 35$^{\circ}$, -1.5$^{\circ}< \ b \ <$ 1.5$^{\circ}$, up to redshift $z <$ 4500 \kms. %We didn't find any conclusive evidence that \HI mass distribution varies in the Void, but  
We found that the \HI mass and LV–centric radius are weakly correlated, but with a high significance. However, number statistics in voids are of course very low.

We find  a total of 9 group candidates two of which (one with 5 members and one with 3 members) are inside the Void. Three groups (with 8, 4, and 3 members respectively) form part of the border of Void and four of them (one with 6, three with 3 members) are in average density regions. Some groups in and at the edge of the Void are consistent with galaxies lying along intra-void filaments, which may suggest the ongoing growth of these galaxies along a filament inside the Void. 

These results are consistent with other studies that observe voids containing a population of galaxies some of which present evidence for ongoing gas accretion, interactions with small companions and filamentary alignments \citep[e.g.][]{kreckel11, kreckel12, beygu13, chengalur13, chengalur15, chengalur17, kurapati20}. The small-scale clustering of the \HI selected galaxies at the centre of the Void is higher than that of clustering in outskirts of the Void and  the field excluding the GA wall, but is lower than clustering in relatively overdense GA wall. This is in agreement with \citet{abbas07}, who found that the galaxies in least dense regions are more clustered than galaxies in moderately underdensities, but galaxies in moderate underdensities are less clustered than those at moderate overdensities.

\section*{Acknowledgements}

 We thank the anonymous referee for insightful comments that improved this paper. SK, RKK, and DJP greatly acknowledge support from the South African Research Chairs Initiative of the Department of Science and Technology and National Research Foundation.

 The MeerKAT telescope is operated by the South African Radio Astronomy Observatory, which is a facility of the National Research Foundation, an agency of the Department of Science and Innovation. We acknowledge the use of the ilifu cloud computing facility—www.ilifu.ac.za, a partnership between the University of Cape Town, the University of the Western Cape, the University of Stellenbosch, Sol Plaatje University, the Cape Peninsula University of Technology, and the South African Radio Astronomy Observatory. The Ilifu facility is supported by contributions from the Inter-University Institute for Data Intensive Astronomy (IDIA a partnership between the University of Cape Town, the University of Pretoria, the University of the Western Cape and the South African Radio Astronomy Observatory), the Computational Biology division at UCT and the Data Intensive Research Initiative of South Africa (DIRISA).

\section*{Data Availability}
The full galaxy catalogue and the atlas are available online. The ASCII file of the GPS catalogue will be submitted to SIMBAD. The raw MeerKAT data are archived and are available at \url{https://archive.sarao.ac.za}. Reduced data cubes will be available on request.

\section*{Supporting Information}
 Supplementary data are available online at MNRAS. The full LV-SMGPS catalogue is provided in machine-readable format as Catalogue.txt, and the complete LV-SMGPS atlas is accessible as atlas.pdf.
%%%%%%%%%%%%%%%%%%%%%%%%%%%%%%%%%%%%%%%%%%%%%%%%%

%%%%%%%%%%%%%%%%%%%% REFERENCES %%%%%%%%%%%%%%%%%%

% The best way to enter references is to use BibTeX:

\bibliographystyle{mnras}
\bibliography{Local_void} % if your bibtex file is called example.bib

% Alternatively you could enter them by hand, like this:
% This method is tedious and prone to error if you have lots of references

%\end{thebibliography}

%%%%%%%%%%%%%%%%%%%%%%%%%%%%%%%%%%%%%%%%%%%%%%%%%%

%%%%%%%%%%%%%%%%% APPENDICES %%%%%%%%%%%%%%%%%%%%%
%\clearpage
\appendix
%\section{LV--SMGPS Catalogue}
%\label{LV_catalogue}

%In this appendix, we present the full LV--SMGPS catalogue. Table \ref{table:app1}  lists all the \HI parameters for the detected galaxies.

%The columns are as follows: \\
%{\it Column 1} - Name of the galaxy: The format (SMGPS--HI) J{\small HHMMSS-DDMMSS} represents the rounded values of RA and Declination,  \\
%{\it Column 2} -  Mosaic (Txx or Nxx from Figure \ref{fig:mosaic}) in which the galaxy is detected, \\
%{\it Column 3} -  Galactic longitude, $\ell$ in deg, \\
%{\it Column 4} -  Galactic latitude, $b$  in deg, \\
%{\it Column 5 and 6 } -  Integrated flux, $S_{\rm int}$ along with its corresponding error $\sigma_{S_{\rm int}}$ in \Jykms,  \\
%{\it Column 7} -  Peak flux, $S_{\rm peak}$ in Jy, \\
%{\it Column 8} -  Measured local RMS around the detection in \mJybeam  per 44.1 \kms channel, \\
%{\it Column 9} - Optical heliocentric velocity, $V_{\rm hel}$ in \kms, \\
%{\it Column 10 and 11} - Line widths $w_{\rm 50}$ and $w_{\rm 20}$ in km/s. These values represent the measured linewidths at 50 $\%$ and 20 $\%$ of the peak flux, respectively. \\
%{\it Column 12} -  Logarithm of the \HI mass, log $M_{\rm HI}$ in log (\msun) \\
%{\it Column 13} - Flag (category 1 indicates a definite detection and category 2 indicates a possible detection), \\
%{\it Column 14} -  Notes on galaxies. It describes whether a galaxy is interacting, has a companion, or any other relevant information.

\section{LV-SMGPS Atlas}

\begin{figure*}
\centering
\includegraphics[width=0.49\linewidth]{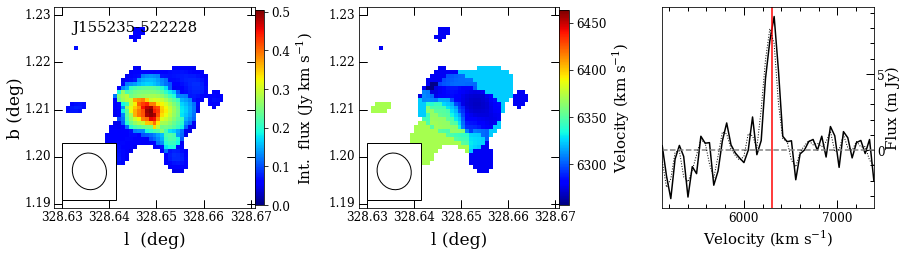}
\includegraphics[width=0.49\linewidth]{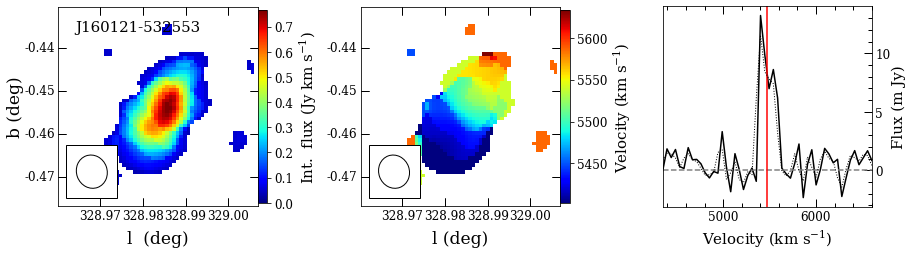} \\
\includegraphics[width=0.49\linewidth]{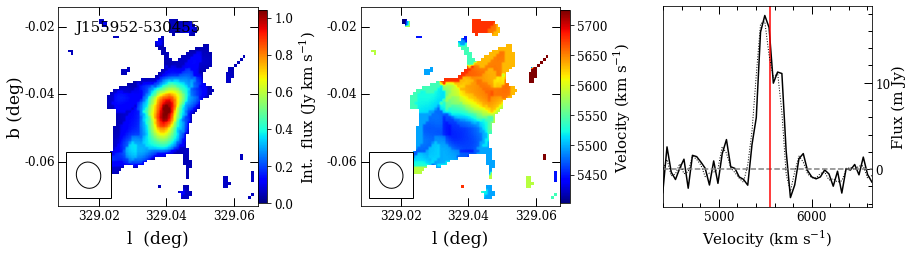}
\includegraphics[width=0.49\linewidth]{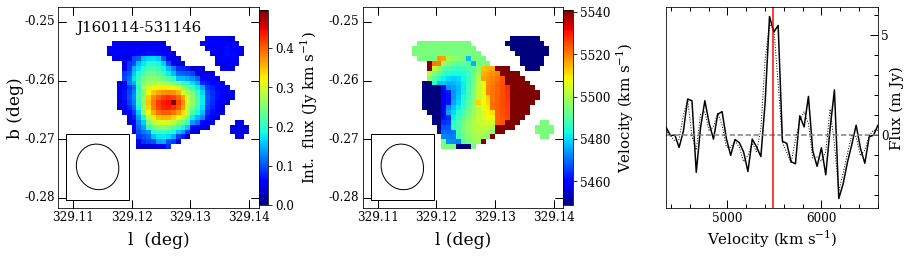}\\
\includegraphics[width=0.49\linewidth]{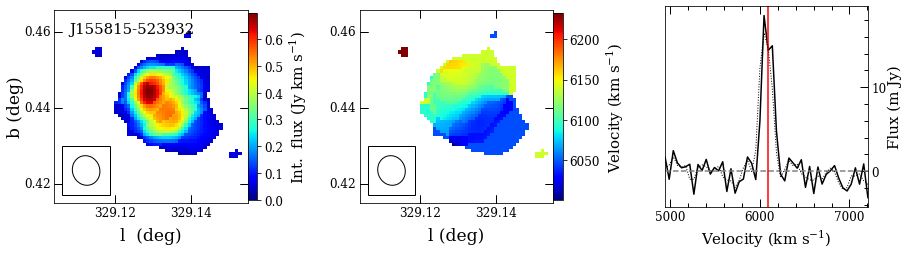}
\includegraphics[width=0.49\linewidth]{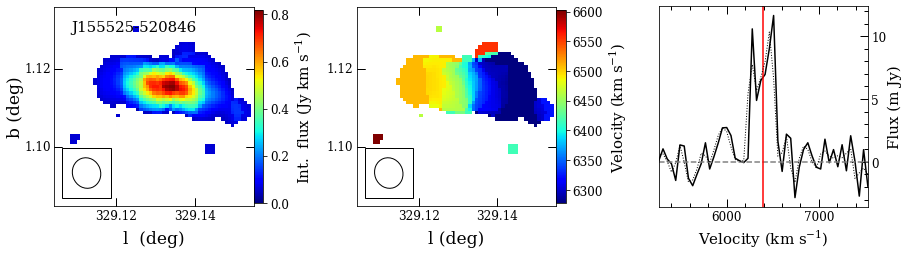} \\
\includegraphics[width=0.49\linewidth]{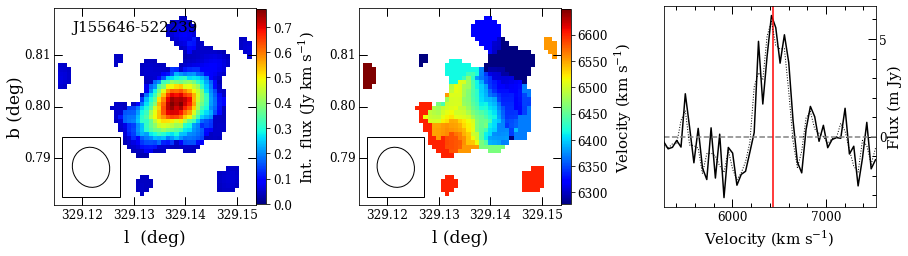} 
\includegraphics[width=0.49\linewidth]{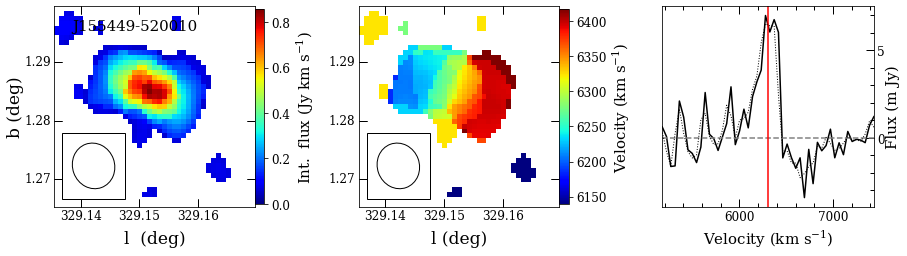} \\
\includegraphics[width=0.49\linewidth]{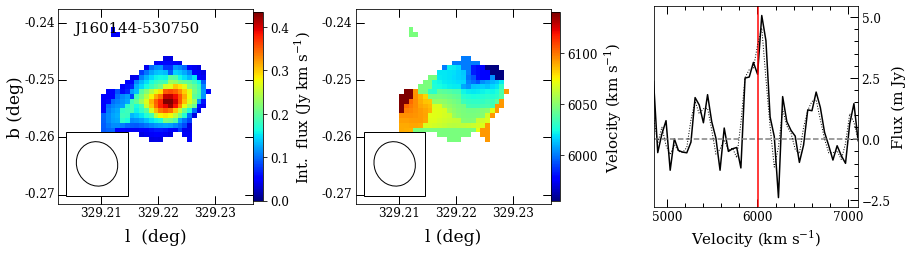} 
\includegraphics[width=0.49\linewidth]{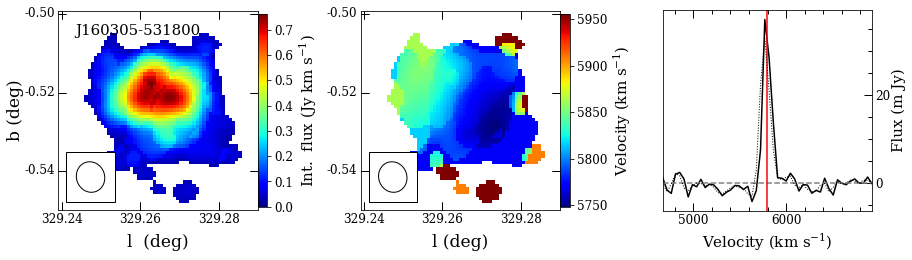} \\
\includegraphics[width=0.49\linewidth]{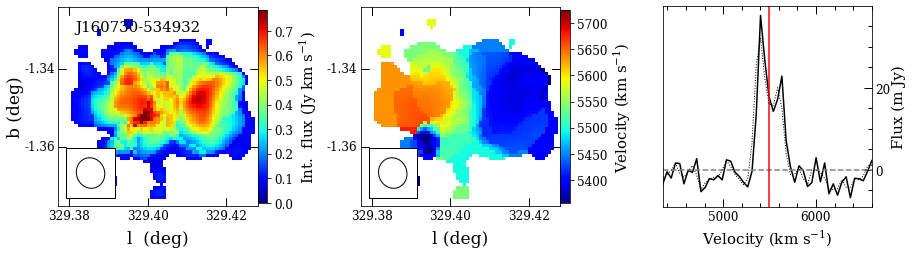}
\includegraphics[width=0.49\linewidth]{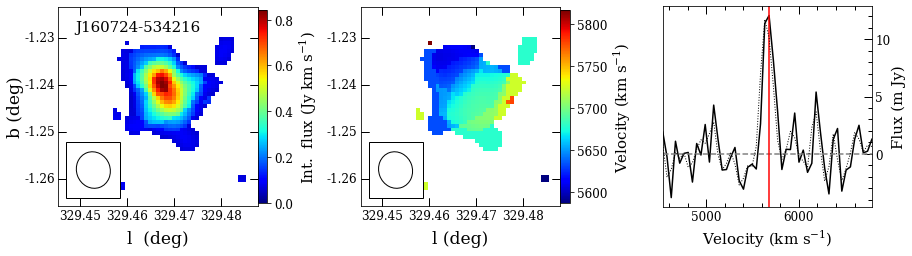} \\
\includegraphics[width=0.49\linewidth]{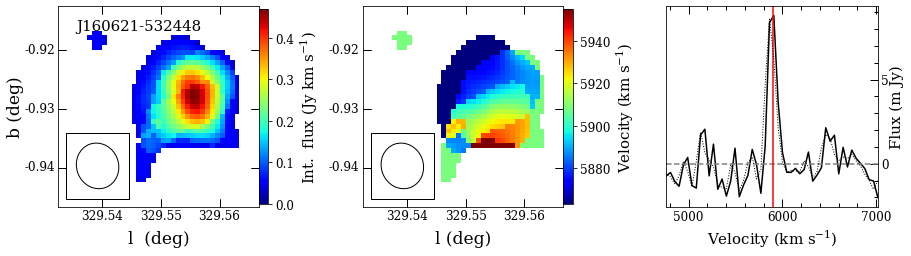} 
\includegraphics[width=0.49\linewidth]{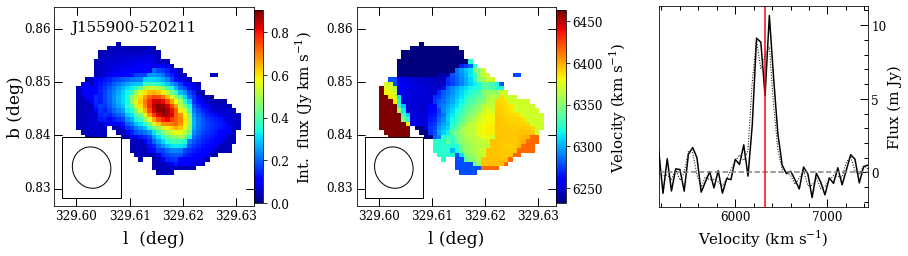} \\
\includegraphics[width=0.49\linewidth]{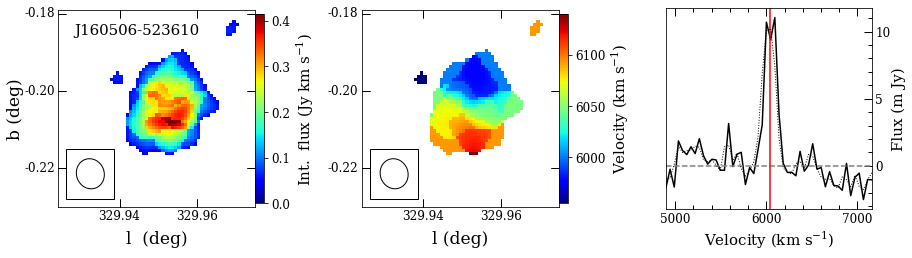} 
\includegraphics[width=0.49\linewidth]{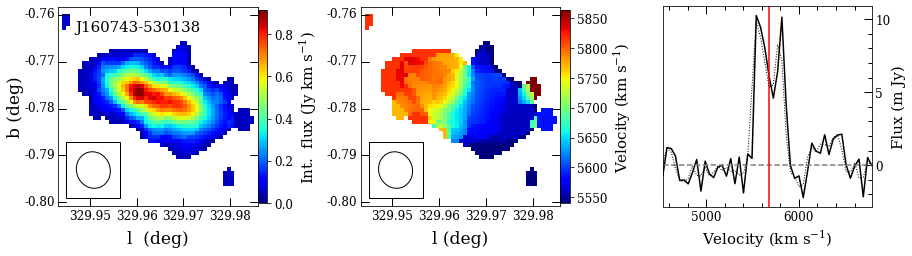} \\
\caption{Moment maps and spectra for a few of the detected galaxies are shown in this figure. The moment maps and spectra of all detections are available online.}
\label{fig:mom1}
\end{figure*}

In this appendix, we present moment maps and spectra for some of the detected galaxies. Figure \ref{fig:mom1} displays: (i) The integrated H{\sc i} intensity map. The galaxy ID (Jhhmmss $\pm$ ddmmss) for each galaxy is shown on the top-left of moment 0 map.  (ii) The intensity-weighted first moment of the galaxy, The synthesis beam FWHM is indicated by an ellipse in both the moment 0 and moment 1 maps. and (iii) The global \HI profile of the galaxy is shown by solid black line. The smoothed profile by a black dashed line and the solid red line indicates the heliocentric velocity. The moment maps and spectra for all detected galaxies are available online.
%put Flagn plots here and Flag2 plots at the end.

% Don't change these lines
%\bsp	% typesetting comment
\label{lastpage}
\end{document}